\documentclass[fleqn,usenatbib,useAMS]{mnras}

\usepackage{graphicx}	
\usepackage{environ}
\usepackage{amsmath}	
\usepackage{amssymb}	
\usepackage{multicol}        
\usepackage{bm}		
\usepackage{pdflscape}	
\usepackage{relsize}
\usepackage{capt-of}

\newcommand{\Msun}{M$_\odot$}

\newcommand{\MUV}{M_{\mathrm{UV}}}

\newcommand{\phistar}{\phi^*}
\newcommand{\Mstar}{M^*}
\newcommand{\vulcan}{\textsc{Vulcan} }
\newcommand{\fesc}{f_{\mathrm{esc}}}
\NewEnviron{myequation}{%
    \begin{equation}
    \scalebox{1.5}{$\BODY$}
    \end{equation}
    }

\usepackage[T1]{fontenc}
\usepackage{ae,aecompl}
\usepackage{comment}
\usepackage{txfonts}

\title[The Little Galaxies that Could]{The Little Galaxies that Could \textit{(Reionize the Universe): \\ Predicting Faint End Slopes \& Escape Fractions at z$>$4}}

\author[L. Anderson]{Lauren\ Anderson$^{1}$, F.\  Governato$^{1}$, M. Karcher$^{1}$, T.\ Quinn$^{1}$ and J.\ Wadsley$^{2}$\thanks{Contact e-mail: \href{mailto:lmanders@uw.edu}{lmanders@uw.edu}}\thanks{Present address: Astronomy Department, University of Washington, \mbox{Box 351580} U.W. Seattle, WA 98195-1580}
\\
$^{1}$Department of Astronomy, University of Washington, Seattle, WA \\
$^{2}$Department of Physics and Astronomy, McMaster University, Hamilton, Canada}
\date{Last updated 2015 Aug 15; in original form 2015 Aug 5}

\pubyear{2016}

\begin{document}
\maketitle

\begin{abstract}
The sources that reionized the universe are still unknown, but likely candidates are faint but numerous galaxies. In this paper we present results from running a high resolution, uniform volume simulation, the \vulcan, to predict the number densities of undetectable, faint galaxies and their escape fractions of ionizing radiation, $\fesc$, during reionization. Our approach combines a high spatial resolution, a realistic treatment of feedback and hydro processes, a strict threshold for minimum number of resolution elements per galaxy, and a converged measurement of $\fesc$. We calibrate our physical model using a novel approach to create realistic galaxies at $z=0$, so the simulation is predictive at high redshifts. With this approach we can (1) robustly predict the evolution of the galaxy UV luminosity function at faint magnitudes down to $\MUV\sim-15$, two magnitudes fainter than observations, and (2) estimate $\fesc$ over a large range of galaxy masses based on the detailed stellar and gas distributions in resolved galaxies. We find steep faint end slopes, implying high number densities of faint galaxies, and the dependence of $\fesc$ on the UV magnitude of a galaxy, given by the power-law: $\mathrm{log} \; \fesc = (0.51 \pm 0.04)\MUV + 7.3 \pm 0.8$, with the faint population having $\fesc\sim35\%$. Convolving the UV luminosity function with $\fesc(\MUV)$, we find an ionizing emissivity that is (1) dominated by the faintest galaxies and (2) reionizes the universe at the appropriate rate, consistent with observational constraints of the ionizing emissivity and the optical depth to the decoupling surface $\tau_{es}$, without the need for additional sources of ionizing radiation.
\end{abstract}




\begingroup
\let\clearpage\relax
\endgroup
\newpage


\begin{figure*}
\centering
\includegraphics[trim={4.5cm 0cm 4.5cm 17cm},clip,scale=0.475]{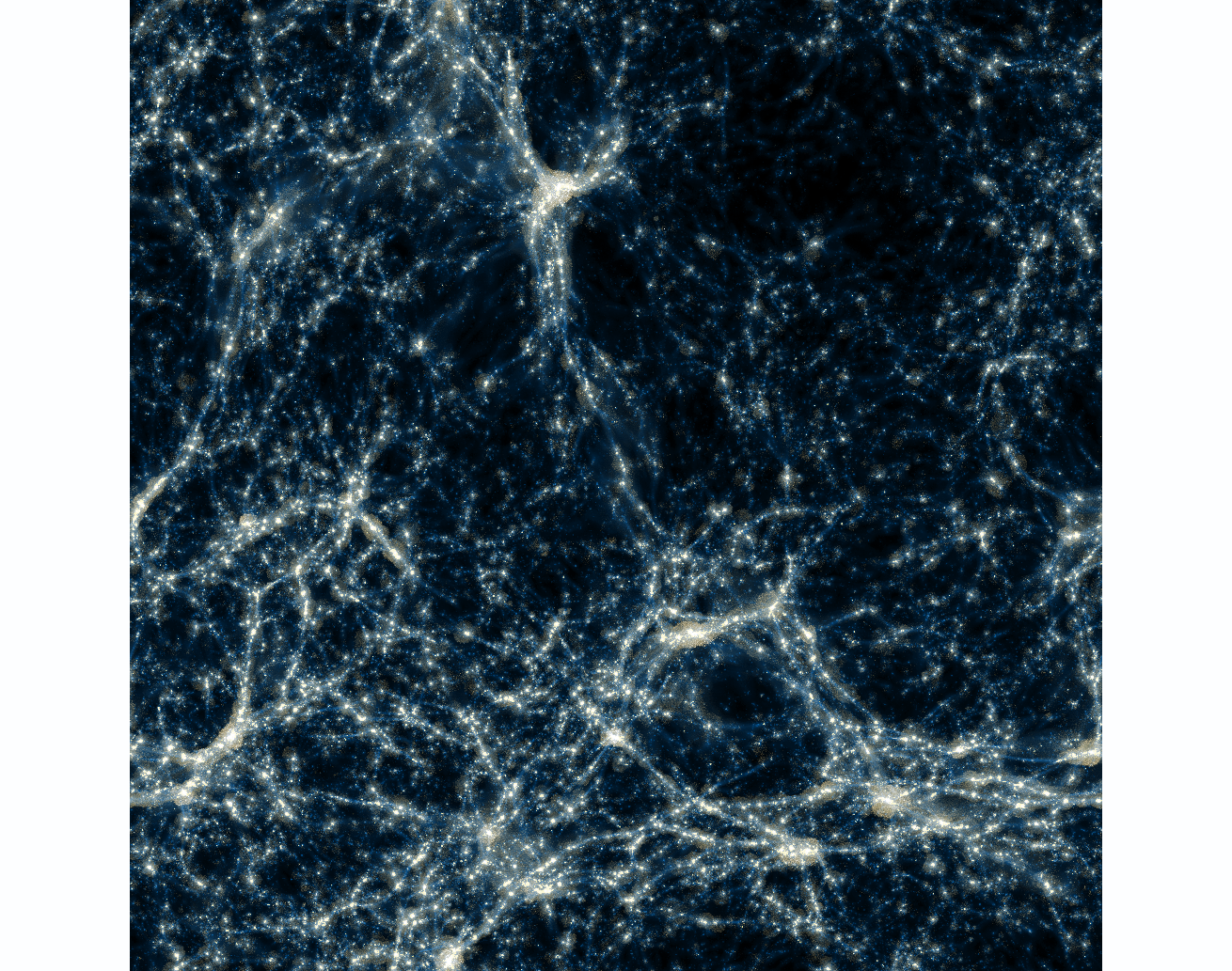}
\caption{A slice of the \vulcan simulation, 25x12x10 Mpc showing the high dynamic range of the simulation at $z\sim 4$, where the blue hues are gas density, the yellow hues are gas temperature, and the white hues are young stars, less than 50 Myrs old. The gas density traces the filamentary structure of the cosmic web, and the gas temperature traces gas being heated from stellar feedback. With 350 pc resolution in a 25 Mpc box, we resolve the morphologies of $\sim 100$ systems down to $5\times10^9$ \Msun\ in stellar mass, of order the Large Magellanic Cloud, throughout the volume. This gives a statistical sample of lower mass halos with established escape fractions and star formation histories.}

\captionof{table}{The \vulcan Simulation Numerical Parameters}

\begin{tabular}{|l|l|l|l|l|l|l|}
\hline

N Dark & N SPH & M$_{dm}$ [M$_{\odot}$] & M$_{gas}$ [M$_{\odot}$] & M$_{*}$ [M$_{\odot}$] ${}^\ast$ & Softening [pc] ${}^\dagger$ & Run Time [CPU Hours] \\
\hline

$1152^3$ & $768^3$ & $1.22 \times 10^5$ & $2.12 \times 10^5$ & $ 6.36 \times 10^4$ & 311 & 10 million\\

\hline
\multicolumn{7}{|l|}{$\ast$ Mass of the star particles at time of formation.}\\
\multicolumn{7}{|l|}{$\dagger$ Gravitational softening; defined as the half-width of the spline density kernel.}\\
\end{tabular}
\label{viz}
\end{figure*}


\section{Introduction}
Reionization of cosmic hydrogen between $z\sim6-15$ \citep{fan06,bolton13,planck15} was likely caused by the first galaxies, with a significant contribution of ionizing photons coming from the faint, low mass population. Although a likely source, their contribution is difficult to quantify. The largest uncertainties in quantifying their contribution are their photon budget and the efficiency that their photons make it to the intergalactic medium (IGM). The photon budget of faint galaxies is quantified observationally through the slope of the UV luminosity function (UVLF), and the efficiency of escape is quantified by the escape fraction $\fesc$.

Observations are starting to shed light on these uncertainties. HST surveys have constrained the faint end of the UVLF down to $M_{UV} \sim -17 (-19)$ at z $\sim 4 (8)$ \citep{finkelstein14, bouwens15b}, but $\fesc$ at high redshifts is a more difficult measurement. Intervening neutral hydrogen between us and a high redshift galaxy absorbs any ionizing radiation that may have escaped the host halo. In the local universe, low metallicity, high star forming galaxies are the most likely candidate for a significant $\fesc$, but there are very few galaxies with these characteristics, so lower redshift measurements of $\fesc$ are also difficult. Measurements have been attempted on a handful of these local galaxies and show very few ionizing photons escape their host halos \citep{leitet13}.

Even with this progress from observations, uncertainties remain in our understanding of reionization. The faint end slope of the UVLF is a crucial parameter for quantifying the photon budget of faint galaxies. UVLFs that increase steeply at faint magnitudes \citep{grogin11} imply a larger role played by these numerous dwarf galaxies as a source of ionizing photons. Indeed, recent surveys point to slopes steeper than $-2$ at $z > 6$ and to steeper slopes at higher and higher redshifts \citep{finkelstein14}. \footnote{At slopes steeper than $-2$, the galaxy number density formally diverges, however, galaxies fainter than  $\MUV > -6$ are not observed in the Local Universe.} Steeper UVLFs imply a large number of galaxies below the current detection limit of high redshift surveys, but it is uncertain how faint these steep UVLF can be physically extrapolated. A natural turn over in the UVLF is expected due to star formation being inefficient in low mass halos, but the UV magnitude this occurs at has not been observed and is still unknown. So extrapolating the observed faint end slope of the UVLF to magnitudes fainter than observations probe might over predict number of faint galaxies and their ionizing photon budget.

In addition to the uncertainty in the number density of high redshift galaxies, we have very little understanding of their escape fractions as well. Our constraints on $\fesc$ are limited by our understanding of the 3D distribution of gas and stars in high-z star forming regions. Progress is further hampered by a lack of numerical studies of high-z galaxy formation where the baryon distribution is resolved to sub-kpc scales \citep{wise12,ferrara14,cen14}. Resolving the gas and stellar morphology is particularly important since the distribution of gas with respect to stars affects $\fesc$. In particular, more massive systems with disk-like configurations may have low escape fractions \citep{wise09}, while lower mass dwarfs, with bursty SF suggested both by observations and theoretical models \citep{vanderwel11,kauffmann14,G14,yue14}, drive galactic-scale outflows of gas, decreasing the cold gas covering factor and therefore possibly facilitating a higher $\fesc$ \citep{Erb15}.

Since escape fractions at high redshift cannot be directly observed, inferring the contribution of galaxies to reionization relies heavily on simulations. So there has been a lot of work in simulating escape fractions but so far results have suggested the escape fractions of high redshift galaxies are all over the place and therefore still highly uncertain. Some find that the escape fraction tends to increase with halo mass \citep{gnedin08}, others find that the escape fraction decreases with halo mass \citep{yajima11, kimm14, wise14}, and still some find that there is no clear dependence on halo properties \citep{ma15}. These findings tend to have one thing in common: there is large scatter in the values of $\fesc$ for a given halo or halo mass range. These differences might be driven by the fact that each group uses different methods on galaxies with different merger and star formation histories, and most importantly at different resolutions with different subgrid physics implementations.

\begin{figure*}

\centering
\minipage{\textwidth}
\includegraphics[trim={0cm 0cm 0cm 9cm},clip,scale=0.9]{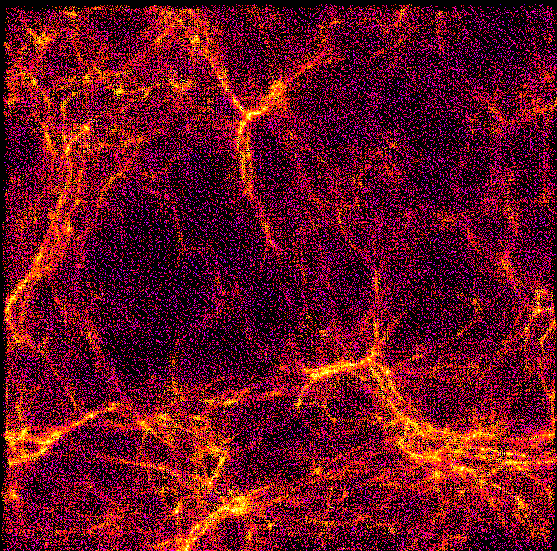}
\endminipage
\caption{{\bf Computational load of the \vulcan simulation}: A visualization of a 25x12x4 Mpc slice of the VULCAN simulation at $z = 3.4$, where each point shows the centroid of a virtual processor.  One million virtual processors are used, each containing $\sim 2000$ particles. The coloring indicates the computational load associated with each virtual processor, where bright, golden pieces have an individually large load and dark, purple pieces have a small load. The load varies by a factor of 8 across the volume and is generally higher within collapsed structure, which is visualized by their bright, gold appearance. The Charm++ runtime system maps these virtual processors onto the real processors of the Blue Waters machine so that the sum of the computational load is approximately equal across all real processors, allowing this highly clustered data set to scale efficiently. With this scaling efficiency for clustered data, we can run a large, uniform cosmological simulation with a high spatial resolution to resolve the internal 3D structure of a statistical sample of faint galaxies.}
\label{load}
\end{figure*}

Substantial progress could then be achieved by simulating a significant number of high-z cosmological systems spanning a range of masses, dynamical states and environments to measure $\fesc$ and determine the faint end slope of the UVLF. These simulations would need to resolve the 3D distribution of stars and gas down to sub-kpc scales, where the properties of the ISM become important. To achieve this, we applied a SF model to a uniform volume cosmological simulation, where the morphology of galaxies down to $10^6$ \Msun\ in stellar mass are resolved down to a nominal force resolution of $\sim 350$  pc. Given that the typical size of such low mass, high-z galaxies is smaller than a few kpc, this is a significant improvement compared to other recent simulations of uniform volumes \citep{illustris14,eagle14}. By resolving the spatial distribution of the gas and stars on this scale, our simulation robustly captures $\fesc$ in smaller, higher redshift galaxies than previous work. An important novelty in our approach is that the SF and feedback parameters were optimized to result in realistic z=0 galaxy models over a wide range of masses, and is described further in Tremmel+16. In addition, we studied in detail the convergence of the escape fraction calculation for our resolved systems. (see resolution study Appendix A).

Our approach combines a high spatial resolution, a realistic treatment of feedback and hydro processes, a strict threshold for minimum number of resolution elements per galaxy, and a converged measurement of $\fesc$. With this approach we (1) {\it robustly predict the evolution of the galaxy LF at faint magnitudes} down to $\MUV \sim -15$ on a sample of galaxies comparable in number to HST data samples and (2) {\it estimate the escape fractions over a large range of galaxy masses based on the detailed spatial distribution of both the stellar and gas components}. Together, this will quantify the photon budget of faint galaxies and their contribution to reionization. The simulations methodology is described in \S2, the results are presented in \S3 and the consequences for reionization are discussed in \S4. We summarize our conclusions in \S5.

\section{The \vulcan Simulation}


To quantify the contribution of faint galaxies to reionization we ran the \vulcan, a high resolution, uniform volume simulation in a full cosmological context to a redshift of 3.4 using the N-body Treecode $+$ Smoothed Particle Hydrodynamics (SPH) code {\sc ChaNGa} \protect\citep{changa08,changa13,menon14}\footnote{ {\sc ChaNGa} is part of the {\sc{AGORA}} group, a research collaboration with the goal of comparing the implementation of hydrodynamics in various cosmological codes \citep{AGORA} on the Blue Waters machine.  {\sc ChaNGa} is available at http://www-hpcc.astro.washington.edu/tools/changa.html}. {\sc ChaNGa} includes standard physics modules previously used in {\sc Gasoline} \citep{wadsley04,shen10} including a treatment of metal line cooling, self shielding, cosmic UV background \citep{HM12}, star formation, turbulent diffusion of metals and thermal energy, and ``blastwave'' Supernovae (SN) feedback \citep{stinson06}. For self shielding, we use the method from \cite{pontzen08} where the ionization state of the hydrogen and helium gas is determined from the cosmic UV backround using an equilibrium solver algorithm based on \cite{katz96}. With the ionization state determined, the attenuation of the UV field due to the H I, He I and He II ions is then calculated. This process is crude compared with radiative transfer techniques, but still important in dense environments.

\begin{figure*}

 \minipage{0.5\textwidth}
  \includegraphics[trim=0mm 4mm 0mm 4mm, clip=true, width=8.75cm]{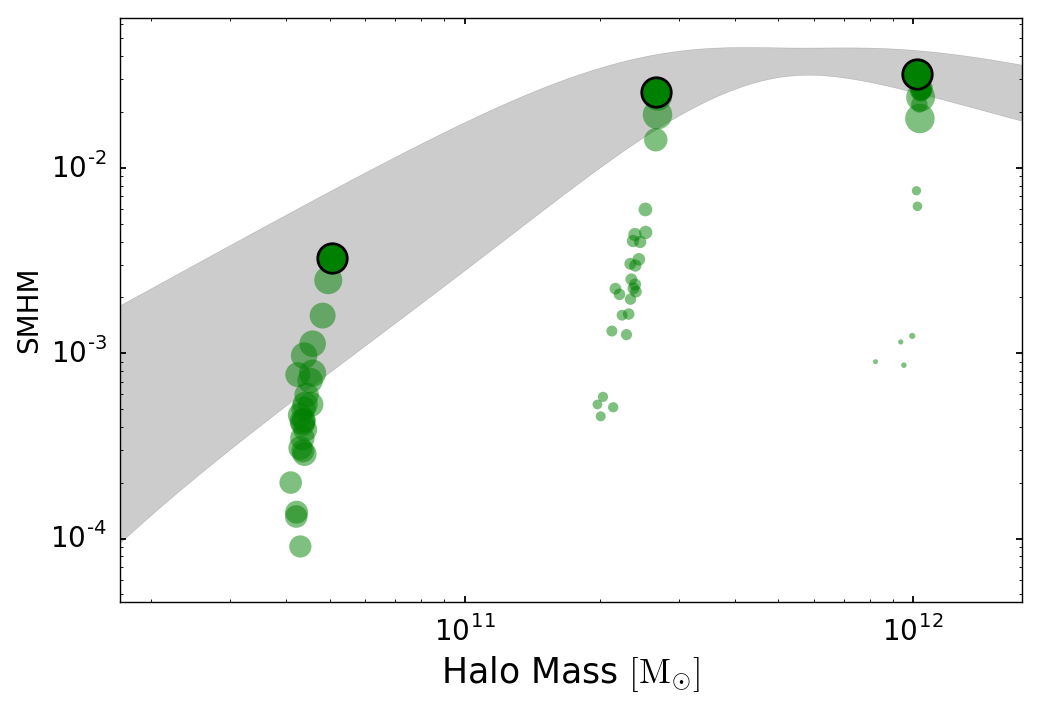}
  \endminipage
  \minipage{0.5\textwidth}
  \includegraphics[trim=0mm 0mm 0mm 0mm, clip=true, width=8.75cm]{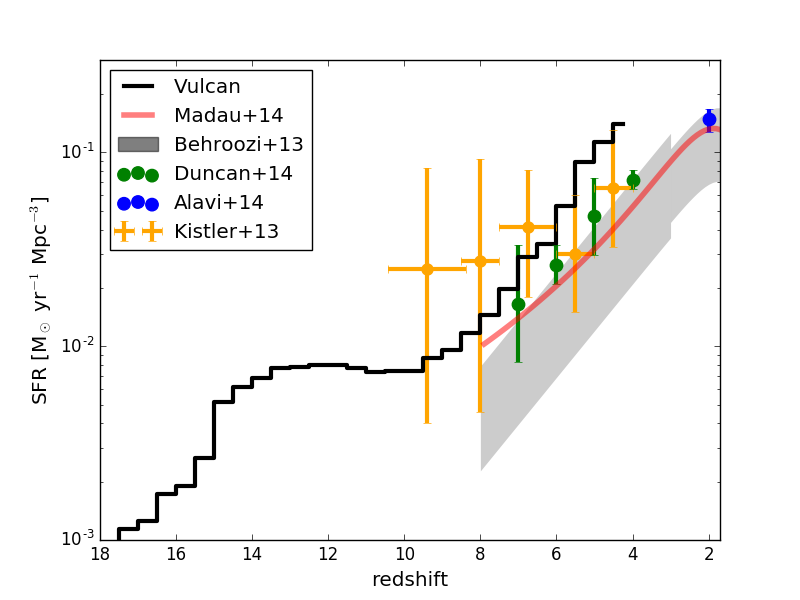}
  \endminipage
\caption{{\sc Left:} {\bf The parameter search}: A comparison of the the observed stellar mass-halo mass (SMHM) relation with the SMHM ratio measured from zoomed-in simulations, all at $z=0$.
The grey region represents the observed SMHM relation with its uncertainties.
Each green point represents a realization of a parameter set applied to 3 zoomed in simulations of varying halo masses.
The size of the point represents how well it fits {\bf all} of our observational criteria, described further in the text.
The best set is outlined in black.
This parameter set simulates galaxies that match the SMHM relationship  at all mass scales, as shown in this figure, and simulates galaxies that match the other observational criteria as well.
We adopt this parameter set for the \vulcan simulation, which optimizes the subgrid SF parameters to create realistic galaxies at $z = 0$ so we are predictive at $z = 4-10$.
{\sc Right:} {\bf The \vulcan Cosmic Star Formation Rate} compared with observational probes: The black line represents all stars formed in the \vulcan over cosmic time.
The grey band represents an analytic fit to previous measurements of the CSFR using various observed, rest wavelengths spanning the 6 years up to 2013 \protect\citep{behroozi13b}.
The error bars represent the intrinsic scatter in the set of measurements they extracted from the literature. The red band is a more recent constraint of the CSFR from \protect\cite{madau14c}.
The addition of more recent measurements, shown as individual colored data points, shows a trend of increasing the measured CSFR in agreement with \protect\cite{madau14c} due to surveys probing to fainter galaxies.
And by getting a more complete census of the CSFR, it approaches the \vulcan results, confirming that we have realistic SFRs in our simulated galaxy population.}
\label{paramsearch}
\end{figure*}
\begin{figure*}

 \minipage{0.5\textwidth}
  \includegraphics[trim=0mm 4mm 0mm 4mm, clip=true, width=8.75cm]{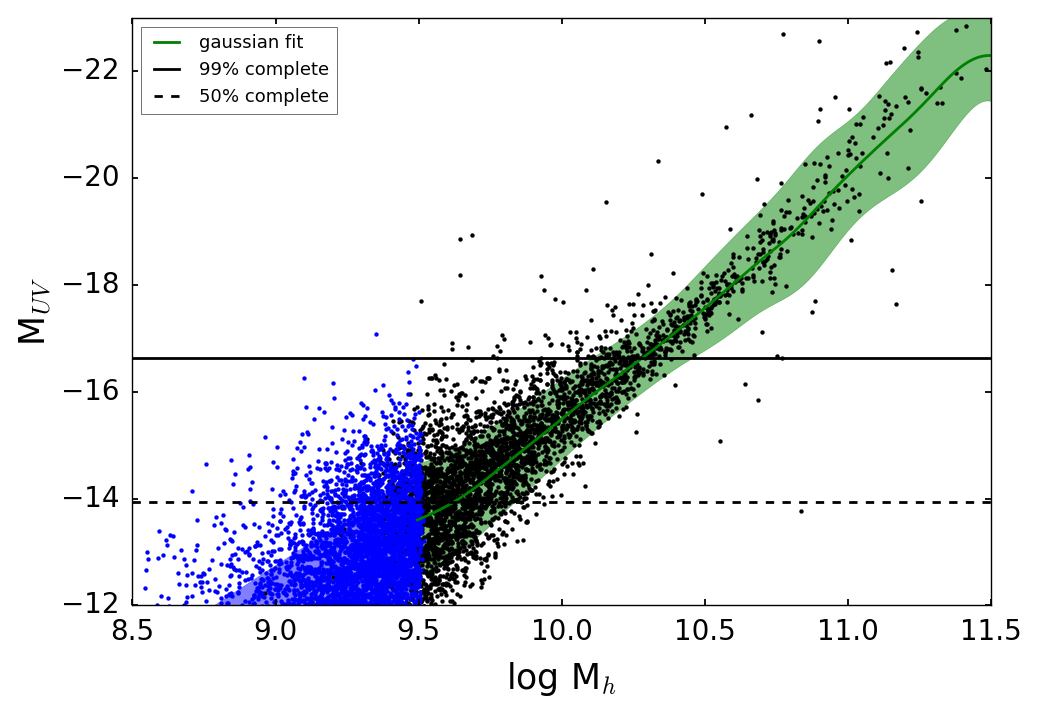}
  \endminipage
  \minipage{0.5\textwidth}
  \includegraphics[trim=0mm 0mm 0mm 0mm, clip=true, width=8.75cm]{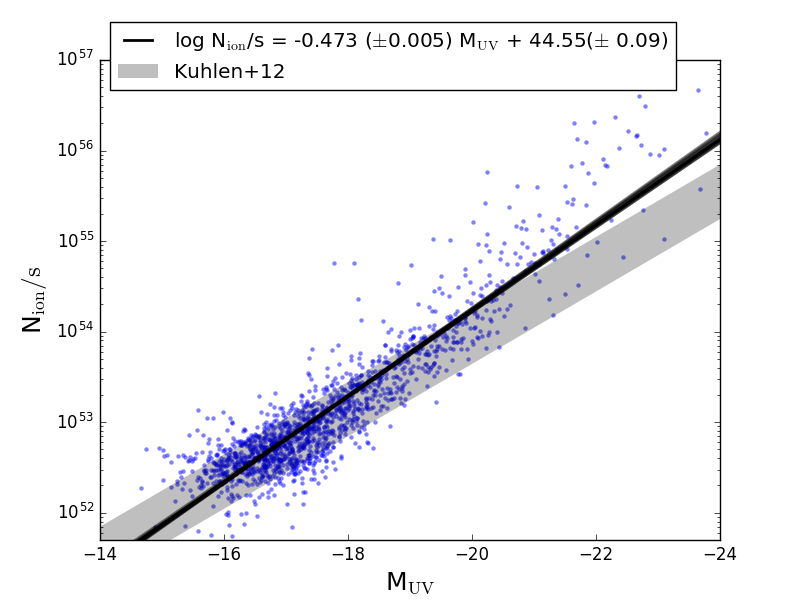}
  \endminipage
\caption{{\sc Left:} {\bf Quantifying Completeness}: The z=4 projection of the $\MUV$ - $M_{halo}$ relationship. Black points are simulated galaxies with reliable star formation histories at z=4, the green band is the 1-$\sigma$Gaussian fit to the kernel density estimation of the black points, and the blue band is the projection of that fit to lower masses. The blue points are a sampling of the projected distribution by the simulated halo mass function of non dark halos. The black line represents the magnitude of 100\% completeness, and the dashed line represents the magnitude of 50\% completeness. By quantifying the completeness we are able to make more rigorous statements about the UVLF down to fainter magnitudes, here $\MUV$ = -14. {\sc Right:} {\bf Conversion to Ionizing Luminosity}: The best fit conversion between $\MUV$ and ionizing luminosity. Blue points are individual galaxies with reliable star formation histories, and the black line is the best fit power-law to the distribution. The grey band is an analytic conversion from \protect\cite{kuhlen12} which assumes some basic properties about the SED of star forming galaxies.}
\label{completeness}
\end{figure*}


\subsection{The Star Formation Parameter Search}
\begin{figure*}
\centering
\includegraphics[scale=0.75,trim=2mm 0mm 0mm 0mm, clip=true]{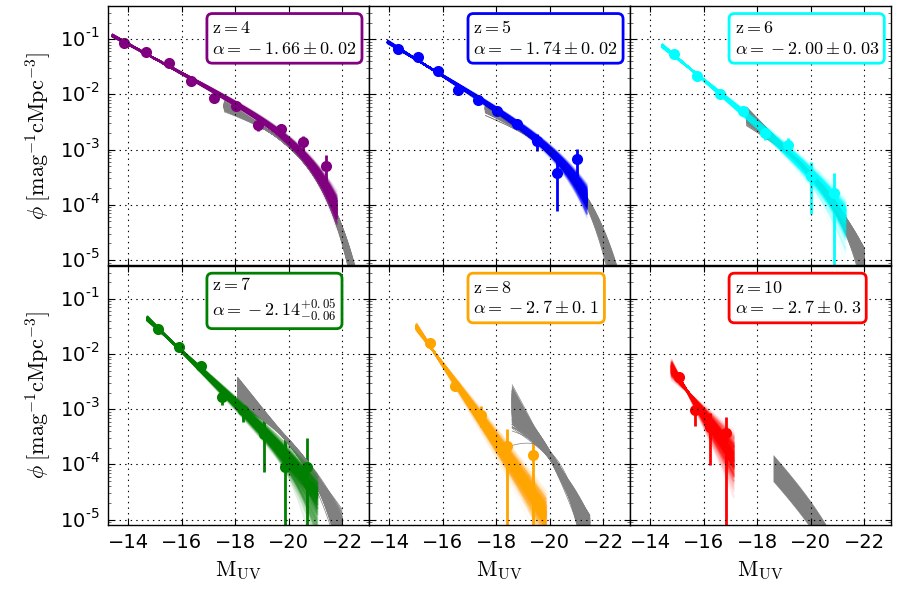}
\caption{The evolution of the simulated UVLF (colored data) from $z \sim 4-10$ compared to observations (grey bands, \protect\cite{finkelstein14,bouwens15b}). Individual colored data points represent the binned simulated UV magnitudes with error bars representing the $1\sigma$ Poisson errors for the number of galaxies in each bin. The colored bands represent the best fit Schecter function to the simulated UV magnitudes using Bayesian inference and sampling the Schecter parameter posteriors. The simulated UVLFs match observations well at the bright end where observations are complete, and constrain the faint end $\sim 2$ magnitudes fainter than observations and predict steeper faint end slopes. The steeper faint end slopes are due to a higher density of faint galaxies below the current detection limits of high redshift surveys.}
\label{uvlf}
\end{figure*}

We calibrated our subgrid star formation parameters to create realistic galaxies at $z=0$. While the parameter space related to SF and feedback in Semi Analytical Models (SAM) can be rapidly explored using Bayesian techniques \citep{benson10a,bower2010}, large, uniform volume simulations are limited by the small number of experiments possible. In this work we adopted a novel approach to validate our model at $z=0$ so we can be predictive at $z \sim 4-10$, during reionization. This approach runs a suite of ``zoomed-in'' simulations to $z=0$ with a strategic variation in the subgrid star formation parameters. We use Gaussian process regression to map out the suitability of the subgrid parameter space for our star formation prescription (Tremmel+16). In our tests we focused on three parameters: (1) the normalization of the SF efficiency, c$^*$, which dictates the fraction of gas that turns into stars in a dynamical time, (2) the amount of energy per SN that is coupled to the ISM, dESN, and (3) the density threshold at which cold gas is allowed to form stars, $\rho_{crit}$. Each set of parameters was applied to three ``zoomed-in'' simulations with masses $\sim$ 10$^{10}$, 10$^{11}$ and 10$^{12}$ \Msun\ at $z = 0$. Each run uses the same cosmology, particle mass, force softening and all numerical parameters as the \vulcan simulation described below, and documented in Table 1. We then measured a number of observable properties for each galaxy and compared each set of three ``zoomed-in'' simulations with the observed real-world scaling relations. We used the cold gas vs stellar mass ratios, the specific stellar angular momentum vs stellar mass \citep{behroozi12,obreschkow14}, and the stellar mass/halo mass ratios \citep{behroozi13} with the corrections to stellar and halo mass described in \cite{munshi13}, as shown in Figure \ref{paramsearch}, left. The figure shows the realization of the SMHM relation in green for each simulated galaxy at $z=0$ for all of the parameter options chosen, 20 for each halo mass. The grey band is the observed relationship and its uncertainties from \cite{guo10}. The black circled points represent the realization for the parameters we chose to use in the \vulcan simulation. This underlying parameter choice not only matches the observed SMHM well, as shown in Figure \ref{paramsearch}, left, but also matches the other observed constraints as well. Each simulated galaxy at $z=0$ received a ``grade'' based on its distance from the observed relation, and the Gaussian process algorithm penalizes parameter values that lead to simulations that deviate from the properties of real, local galaxies. It interpolates between the tested values to give robust estimates of uncertainty around parameter values we have not yet simulated. At each step we then choose our next parameter values to test based on how likely it is to turn out better than our best values so far.  We then repeat the runs with the same galaxy set, but with the updated parameters, for a number of times until the parameters converge to a set of values that form the most realistic galaxies. Each galaxy was equally weighted, but we explore different choices in Tremmel+16, where dwarfs were more heavily weighted. This approach maximizes the ``goodness'' of our adopted star formation and feedback implementation. In practice, this approach is relatively free of human biases and superior to a standard random-walk Markov chain when only a limited number of experiments can be carried out. For example, we ran 20 sets of three simulations each. The following parameters created the most realistic galaxies across the three mass ranges: (1) c$^* = 0.1$, (2) dESN$ = 10^{51}$ ergs, and (3) $\rho_{crit} = 0.1$ amu/cc. These parameters were constrained from the z=0 observational constraints alone, instead of an evolution due to the z=0 constraints being much stronger, especially for morphology. In addition, by using constraints from $z=0$, rather than from high redshift, our simulations are {\it predictive} at high redshift, and therefore able to predict the faint end of the UVLF at high z. As independent validation, our simulated UVLF independently matching the observed high-z UVLF, see \S3 and Figure \ref{uvlf}, suggests that this method successfully maps out parameter space to find parameters that simulate realistic SFRs and metallicities of galaxies.



\subsection{The Simulation Parameters}
After optimizing the SF parameters, we then ran our main simulation: a uniform volume, high resolution cosmological simulation. We chose a box of 25 comoving Mpc (cMpc) to sample the smaller $k$ modes of the matter power spectrum and get a maximally diverse sample of environments. We also wanted a high enough resolution to resolve the morphologies of low mass galaxies, and to separate the disk and spheroidal kinematic components. With this choice of box size, however, we do not sample the knee of the UVLF at $z > 5$, so to fit the UVLF we take the observed values as a prior to our UVLF parameter search \citep{finkelstein14}, which we'll discuss more in \S3.1. The starting z was 109, and we ran the simulation to $z = 3.4$. Simulating a 25 cMpc box with a spline kernel resolution of 350 pc was made possible due to the strong scaling of ChaNGa, with 50\% parallel efficiency up to {\it 128K cores}. The strong scaling is due in part to the adaptive load balancing strategies of the Charm++ run time system, visualized in Figure \ref{load}. In the visualization, each "particle" is a tree piece representing $\sim 2000$ particles. The coloring indicates the computational load associated with each virtual processor, where bright, golden pieces have an individually large load and dark, purple pieces have a small load. The load varies by a factor of 8 across the volume and is generally higher within collapsed structures, which is visualized by their bright, gold appearance. The Charm++ runtime system maps these virtual processors onto the real processors of the Blue Waters machine so that the sum of the computational load is approximately equal across all real processors, allowing this highly clustered data set to scale efficiently.

We make various optimal parameter choices for the physical modules within the simulation. We assume a Kroupa 2001 IMF \citep{kroupa2001}, which sets the stellar feedback and SN rates, and the density threshold for star formation ($\rho_{crit}$ above) is 0.1 amu/cm$^3$. Limiting star formation to dense gas regions is a realistic approach and concentrates feedback energy to effectively regulate star formation \citep{governato10,brook12,agertz14,keller14}. For each SN event, $10^{51}$ ergs of energy ($dESN$ above) couples to the surrounding gas.  This accounts for energy injected by high mass stars and the subsequent SN. Cooling shuts off in gas receiving SNe energy according to the ``blastwave'' approximation \citep{stinson06}. {\sc ChaNGa} uses a spline kernel gravitational softening \citep{treesph89} so gravity becomes Newtonian at two softening lengths. Periodic boundary conditions are handled by an Ewald summation of the 4th order multipole expansion of the mass distribution in the volume \citep{Stadel01}. The SPH implementation includes thermal and metal diffusion \citep{shen10} and eliminates artificial gas surface tension through the use of a geometric density mean in the SPH force expression \citep{ritchie01,G14,keller14}. The simulation took 10 million CPU hours (see Table \ref{viz}) on Blue Waters. Halos are identified using {\tt ROCKSTAR} \citep{rockstar}, a phase space halo finder. Galaxy magnitudes are computed using SSP model magnitudes assuming a Kroupa 2001 IMF, and reddening is estimated following observational prescriptions described in \S3.1. Finally the escape fractions of individual galaxies are calculated for each system by measuring the covering fraction of HI at resolved scales, described in \S3.3.

\subsection{Star Formation History}
\begin{figure*}
\centering
\includegraphics[scale=0.5,trim=2mm 0.6cm 0mm 0mm,clip=True]{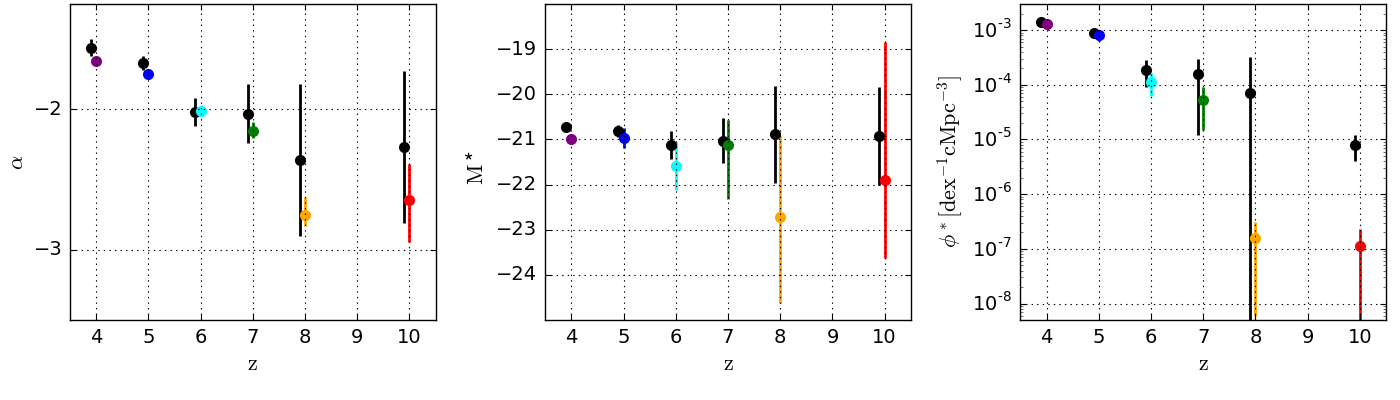}
\caption{Evolution of the Schechter parameters $\alpha$, $\Mstar$ and $\phi^{*}$, in color, compared with observations, in black \protect\citep{finkelstein14,bouwens15b}, from $z \sim 4-10$. The error bars represent $1\sigma$ confidence ranges from the Bayesian inference posteriors. The faint end slope, $\alpha$, is consistently steeper than observations, but within the observational uncertainties. The exponential cut off, $\Mstar$, is pushing to the brighter variance of the observations, with large uncertainties. Our large uncertainties in $\Mstar$ are a reflection of our box size which does not fully sample the bright end of the luminosity function at high redshifts, and therefore also coinciding with the brightest observed value. A brighter $\Mstar$ then drives the discrepancy in $\phi^{*}$, the normalization of the UVLF. The simulated $\phi^{*}$, agrees well with the observations except at the highest redshifts, though is still within the uncertainties at $z\sim8$. Overall the simulation is a good match to the data, with the simulated box size dominating the errors at the bright end. The steepening of $\alpha$ is due to a relatively higher density of low luminosity galaxies at high redshifts.}
\label{paramEv}
\end{figure*}

The cosmic star formation rate (CSFR) is a census of stars created over time in an average patch of the universe. High redshift estimates of the CSFR are based on the integration of the observed UVLF down the faint end slope to the limiting magnitude of the survey, usually $\MUV \sim -17$, but not to the turn over in the luminosity function where star formation becomes inefficient. The UV energy density can be converted to a SFR density because the UV light is dominated by young, recently formed stars. Dust absorbs some of the UV light and re-emits the energy in the infrared, a difficult rest wavelength to observe at high redshifts due to a lack of observatories, and therefore hard to correct for. Previous measurements assumed a fixed dust extinction at $z > 2.5$ \citep{hopkins06}. This over corrected for dust \citep{bouwens12,reddy09} and therefore over predicted the UV luminosity density and CSFR. Similarly, current measurements may under predict the steepness of the faint end slope of the UVLF, or the limiting magnitude may be too shallow, and therefore miss a large population of faint galaxies. This would imply a smaller UV luminosity density and therefore under predict the CSFR. For comparison, the CSFR for the \vulcan is shown in Figure \ref{paramsearch}, right, represented by the black histogram, and is a census of {\it all} stars formed in the volume. An analytical fit to all observed measurements of the CSFR up to 2013 \mbox{\citep{behroozi13b}} is shown as the grey band, with error bars representing the scatter in the accumulated measurements from the literature. The red band is a measurements of the CSFR from \cite{madau14c}. The additional individual, colored points represent the CSFR measured more recently than 2013, so not included in the comprehensive literature search represented by the red line. These more recent measurements include star formation from fainter galaxies as surveys probe to fainter magnitudes. In particular, \citep{alavi14} (the blue point) is the inferred CSFR from measuring the $z\sim2$ UVLF to 100x dimmer galaxies than previous surveys, increasing the measured CSFR. The green point \citep{duncan14} represents results from integrating the stellar mass functions from the CANDLES survey which probed to lower stellar masses than previous surveys. The yellow points \citep{kistler13} is the inferred CSFR from observed gamma ray bursts, which although highly uncertain can probe to higher redshifts. Although not directly comparable to our simulation since it is at $z\sim2$, it shows the trend that having deeper photometry, and integrating farther down the UVLF, creates a more detailed picture of a higher CSFR. By including fainter galaxies and getting a better census of star formation, these measurements of the CSFR are larger than the older analytic fit, and are approaching the CSFR of the \vulcan, which includes all the stars formed, even in the faintest halos. This is initial evidence of realistic star formation rates with in each halo, and therefore realistic UVLFs.

\section{Analysis of the \vulcan Simulation}

In this section we analyze the \vulcan simulation to quantify (1) the shape of the UVLF; (2) the conversion between the UV magnitude $\MUV$ and the ionizing photon production rate, $\gamma_{ion}$; and (3) $\fesc$ as a function of $\MUV$. We then use these results to quantify the contribution of faint galaxies to reionization. In section 3.1 we find the Schecter function parameters that best fit the simulated luminosity function including the effects of completeness and dust; in 3.2 we calculate the conversion of $\MUV$ to $\gamma_{ion}$ including the full SFH and stellar metallicities of a halo; and then in 3.3 we measure the escape fractions of the resolved halos, and fit a relation for $\fesc$ as a function of $\MUV$.

\subsection{Measuring the High z UVLF}

{\bf Galaxy Magnitude Calculation:} To make direct comparisons with observations, we measure the rest frame magnitude of our simulated, high redshift galaxies at 1500\AA, in the FUV. To calculate the intrinsic UV magnitude of each galaxy we use {\tt Pynbody} \citep{pynbody}, a lightweight analysis framework for N-body and hydrodynamic astrophysical simulations. The magnitude of each galaxy was calculated using the Padova Simple stellar populations (SSP) from \citep{marigo08, girardi10} \footnote{http://stev.oapd.inaf.it/cgi-bin/cmd}. The SSP model is convolved with the Galex FUV filter, centered on 1525\AA\  with a width of 255\AA, to calculate the 1500\AA\  magnitude. Pynbody then linearly interpolates between a grid of SSP model magnitudes for various stellar ages and metallicities to the desired value for each individual star particle, and sums up the contribution of each star to the total magnitude of the galaxy. The choice of IMF can affect the UV magnitude produced by a galaxy for a given star formation history. If one adopts a top-heavy IMF, the UV magnitude will be brighter and therefor produce more ionizing photons per solar mass created. We assumed a Kroupa 2001 IMF, consistent with the assumed IMF in the \vulcan Simulation.  but adopting a saltpeter IMF would shift the UV luminosity to fainter magnitudes but maintain the same shape, or a charbrier imf would put a slight tilt in the inferred LF compared with Kroupa 2001. This is a post processing choice, but choosing a different IMF in the simulation, and therefore a different number of supernova per stellar mass created would have a nonlinear effect on the SFR and therefore the UVLF.

{\bf Dust Attenuation:} Dust is important when considering the intrinsic versus the observed amount of UV light a galaxy emits. Dust absorbs UV light, processes it, and re-emits it in the infrared. This process dims the observed UV magnitudes of galaxies. Fits to observed luminosity functions do not take this into account when the best Schecter functional fit is modeled. Observers fit directly to the observed data and correct for dust afterwards when quantifying physical quantities, such as the SFR density \citep{smit12}. So to do a proper comparison with observations we dim the modeled $\MUV$ of our simulated galaxies. The effects of dust on low mass galaxies is expected to be minimal, especially at high redshift due to less accumulated metal contamination in the interstellar medium from SNe; however, an attenuation of $\sim 1$  magnitude is expected for the brightest systems \citep{smit12}. To take this into account we use observational estimates of attenuation following the formalism in \cite{smit12}. We direct the reader there for a more thorough description, but summarize their approach here. They use the IRX-$\beta$ relation, which is the correlation the UV-continuum slope, $\beta$, with the ratio of the far infrared flux with the UV flux of a galaxy. This correlation exists due to dust attenuation, which absorbs UV light and reradiates the energy into the infrared, and the most widespread relation is \cite{meurer99} which is a measurement of the correlation from local starburst galaxies. With additional assumptions about the underlying stellar populations and dust attenuation curve, one can relate IRX (and hence $\beta$) to the total FUV dust attenuation. This method of correcting UV slopes for dust attenuation is very common for estimating star formation rates of high redshift galaxies where only measurements of the UV slope $\beta$ are available. One should be careful however, since more recent measurements of the IRX-beta relation have very large scatter, and the deviations from the \cite{meurer99} relation have been observed \citep{conroy13}. Scatter in the intrinsic relationship while using the Meurer relation will cause an incorrect inference for the true amount of star formation in a galaxy, especially since the relationship is a very steep function of $\beta$. There are improvements that can be made here by taking galaxy type into account as well as the observed scatter, but pushing forward they connect the UV slope $\beta$ to the UVLF using the observed correlation of $\beta$ with $\MUV$ in high redshifts samples found in \cite{bouwens12}; for example, at $z=4$, $\beta = -0.11 (M_{UV,AB} + 19.5) - 2.00$. Convolving these two correlations together, they calculate an average attenuation as a function of $\MUV$ and redshift. In future work, we will compare the observed attenuation at $z\sim2$ to the predicted attenuation from our simulations to test if \cite{meurer99} at lower redshifts can be physically applied to higher redshifts.

\begin{figure*}
 \minipage{\textwidth}
\includegraphics[trim=5mm 0mm 0mm 0mm, clip=true, width=17cm]{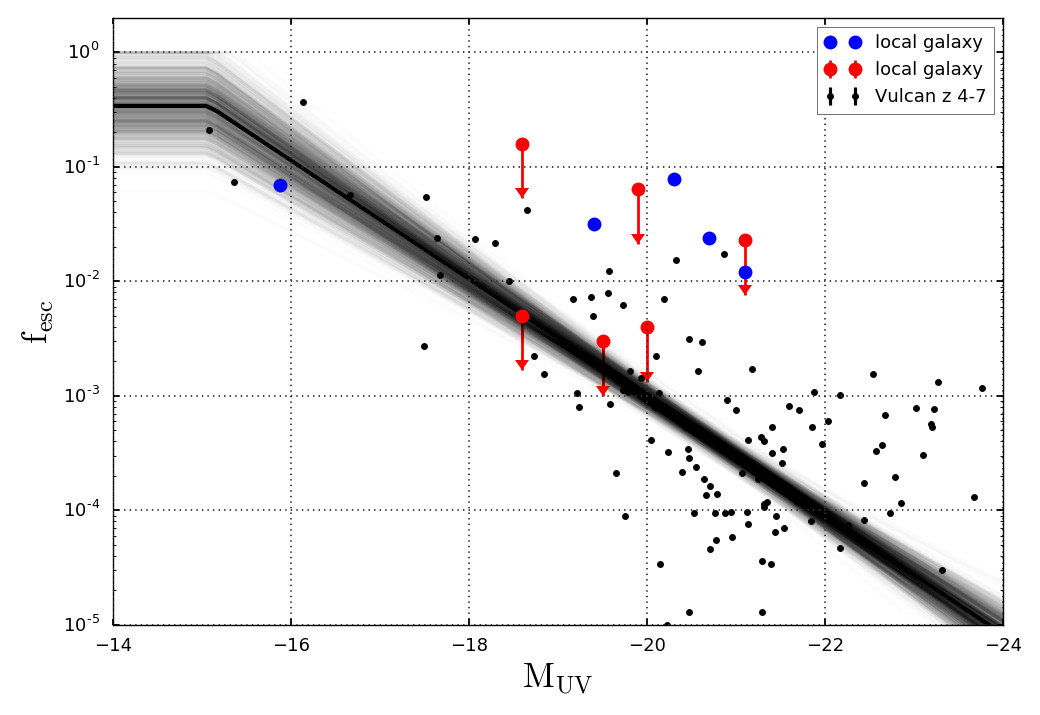}
\endminipage
\caption{Escape fractions of galaxies as a function of their absolute UV magnitudes: {\bf Black points are from our simulation; blue and red points are measurements and upper limits from local galaxies \protect\citep[Choi+ 16]{leitet13}}. The solid black line represents the best fit line $\mathrm{log \ \fesc = (0.51 \pm 0.04) \MUV + 7.3 \pm 0.8}$, and the shaded region shows 1000 samples from the MCMC chain (after the burn in) of the slope and y-intercept from the linear fit. The relationship is capped at the faintest halo that samples the functional fit such that all halos dimmer than $\MUV \sim -15$ have an $\fesc \sim 35$\%. Fainter, low mass halos tend to have higher escape fractions, and brighter, more massive halos have lower escape fractions. This trend agrees with observations of brighter galaxies having little to no escaping ionizing radiation, and supports the theory that faint, low mass galaxies contribute a significant fraction of the ionizing radiation to reionization.}
\label{efMuv}
\end{figure*}

{\bf Completeness:} The uniform volume simulation is a complete sample of halos down to the resolution limit, at which point we must quantify the completeness due to resolution effects. Previous work by \cite{christensen10} investigated the resolution sufficient to establish galactic SFRs, SFHs, stellar feedback, and the distribution of gas and stars. They show that there is a minimum number of particles required to create realistic star formation histories (SFH), and a higher threshold for realistic morphologies. The effects of insufficient resolution can be largely grouped as those relating to overcooling \citep{hutchings00,springel02,governato04}, two-body heating \citep{navarrosteinmetz97,governato07,mayer08}, artificial viscosity \citep{kaufmann07,shen10}, and small-scale gravitational support \citep{bate97}. All of these contribute to unphysically altering positions of the particles and the SFR of the galaxy. Correctly determining the amount and location of SF is vital in simulated galaxies because SF produces and distributes metals throughout the galaxy, affects the distribution of matter in the galaxy through feedback, and enables us to relate simulations to observations.

In \cite{christensen10}, they studied in detail the minimum particle count necessary to simulate realistic SFHs, important for modeling $\MUV$. They also studied the minimum particle count necessary to simulate a realistic morphology, important for modeling $\fesc$, which we will return to in \S3.3. They showed that to simulate realistic galactic SFRs, SFHs, and stellar feedback, a galaxy needs at a minimum $10^4$ dark matter particles and $10^4$ SPH particles. Below this limit, global SFRs and SNe feedback are not accurate, and SFHs, gas loss and accretion are not converged. These galaxy properties are critical to simulating  realistic SFH, and therefore $\MUV$ of a galaxy, so we only calculate $\MUV$ for halos passing this strict resolution threshold.

The shape of the faint end of the simulated UVLF is resolution dependent, and by using a strict resolution threshold our UVLF artificially turns over due to this incompleteness. To quantify the completeness of our sample we use the correlation between $\MUV$ and halo mass, shown in Figure \ref{completeness} left. Brighter galaxies tend live in more massive halos, and the scatter in this relationship grows with decreasing mass. We fit the full distribution of $\MUV$ of as a function of halo mass using a kernel density estimator (KDE). We sample the KDE in constant total mass space at 100 masses down to the $10^4$ particle limit we set, and fit each sample with a Gaussian, shown as the green band in Figure \ref{completeness} left. The minimum particle number for a realistic SFH is high, but we trust the mass accretion of a galaxy, and therefore its halo mass, down to $\sim$ 256 particles. We extrapolate the $\langle\MUV\rangle$ and $\sigma_{\MUV}$ of the Gaussians to lower halo masses, shown as the blue band, sampling the extrapolated Gaussians with the simulated halo mass function, for halos with at least 10 star particles, shown as the blue points. This cut on the number of star particles ensures we were not including dark halos with no star formation in our UVLF calculation, halos that naturally don't have any star formation. Extrapolating down the mass function is important because the spread in magnitudes at a given mass grows with decreasing mass, so brighter, low mass halos can scatter to magnitudes brighter than $\MUV \sim -15$, above our 50\% completeness magnitude. We calculate the completeness of our sample by comparing the halos with reliable SFH to those populated by the projection, as a function of UV magnitude. For the luminosity functions presented below, we fit down to the 50\% completeness level.

{\bf UVLF:} Given our modeled UV magnitudes, we characterize the shape and evolution of the simulated UVLF from $z \sim 4-10$. We fit the UVLF with a Schechter function \protect\citep{schechter}, which matches the observed UVLF from $z\sim0-8$, and is therefore a well motivated functional form to fit. It is characterized by a power-law faint end slope, $\alpha$, an exponential cut-off at the bright end, $\Mstar$, and a normalization $\phistar$. Specifically, $\phistar$ is the Schecter function value at the magnitude $\Mstar$. The number density of galaxies at a given $\MUV$ is given by:
\begin{equation}
\Phi(\MUV) = 0.4 \ln(10)\phistar 10^{0.4(\Mstar -\MUV)(1+\alpha)} \exp\left[-10^{0.4(\Mstar -\MUV)}\right].
\end{equation}

To constrain the free parameters $\alpha$, $\Mstar$, and $\phistar$ we use a Bayesian probabilistic method to calculate the posterior probability distribution. We calculate the probabilities for each parameter by comparing possible Schechter models to the statistical sample of simulated UV magnitudes, and use a Markov chain Monte Carlo (MCMC) sampling algorithm \citep{thehammer} to sample the $\alpha$ and $\Mstar$ parameter space. The prior on $\alpha$ is a top hat spanning (0, -4). The prior on $\Mstar$ is the observed values from \cite{finkelstein14} for z=4-8 and \cite{bouwens15b} for z=10. Specifically it is a Gaussian with a $\mu$ of [-20.73, -20.81, -21.13, -21.03, -20.89, -20.92] and $\sigma$ of [0.09, 0.13, 0.31, 0.50, 1.08, 1.08] for z = 4, 5, 6, 7, 8, 10 respectively. This prior is helpful since the volume does not sample the bright end to constrain this quantity; without this prior the best fit function is a power-law at higher redshifts. Finally, $\phistar$ is then the normalization that constrains the modeled UVLF to create the same number of galaxies as the simulation. This approach allows us to determine the parameters of the UVLF without binning up the data, and therefore captures as much information as possible from each data point.

The evolution of the simulated UVLF is shown in Figure \ref{uvlf}, and the evolution of the best fit Schecter parameters is shown in Figure \ref{paramEv}. In Figure \ref{uvlf}, the colored bands represent the simulated UVLF from $z\sim4-10$ and the grey bands are observations \citep{finkelstein14,bouwens15b}. Individual colored data points represent the binned simulated UV magnitudes with error bars representing the $1\sigma$ Poisson errors for the number of galaxies in each bin. The colored bands represent the best fit Schecter function to the simulated UV magnitudes using Bayesian inference and sampling the Schecter parameter posteriors. The simulated UVLFs match observations well at the bright end where observations are complete, and constrain the faint end $\sim2$ magnitudes fainter than observations and predict steeper faint end slopes. The steeper faint end slopes are due to a higher density of faint galaxies below the current detection limits of high redshift surveys. There is a visible discrepancy in the $z\sim8$ UVLF between the simulation and observations. This is driven by small number statistics. In a 25 Mpc volume it takes about a billion years to grow the galaxies that fill out the bright end of the UVLF, so at $z\sim8$ this is just beginning. The bright bins contain about 1-3 galaxies at this time, until $z\sim6-7$ when the bins become more populated.

The faint end slope of the UVLF is steep at high redshift, and is steeper at earlier times, implying a population of galaxies that is dominated by the faint population at high redshifts. This is demonstrated in Figure \ref{paramEv}, which shows the evolution from $z\sim4-10$ of the Schechter parameters $\alpha$, $\Mstar$ and $\phi^{*}$ as the colored points, compared with observations, in black \protect\citep{finkelstein14,bouwens15b}. The error bars represent $1\sigma$ confidence ranges from the Bayesian inference posteriors. The faint end slope, $\alpha$, is consistently steeper than observations, but within the observational uncertainties. The exponential cut off, $\Mstar$, is pushing to the brighter variance of the observations, with large uncertainties. Our large uncertainties in $\Mstar$ are a reflection of our box size which does not fully sample the bright end of the luminosity function at high redshifts, and therefore also coinciding with the brightest observed value. A brighter $\Mstar$ then drives the discrepancy in $\phi^{*}$, since $\phi^{*}$ is the value of the Schecter function at $\Mstar$. The simulated $\phi^{*}$, agrees well with the observations except at the highest redshifts, though is still within the uncertainties at $z\sim8$. Overall the simulation is a good match to the data, with the simulated box size dominating the errors at the bright end. The steepening of $\alpha$ is due to a relatively higher density of low luminosity galaxies at high redshifts.

\subsection{Conversion from \texorpdfstring{$\MUV$}{Muv} to Ionizing Luminosity}
Measuring the rest frame FUV magnitude of distant galaxies is significantly easier than ionizing emissivity, but the ionizing emissivity is the interesting quantity for reionization. The FUV and ionizing emissivity are both being generated by young, massive stars with well studied spectral energy disrtributions, so we can convert the FUV magnitude of a galaxy to ionizing emissivity. To calculate this conversion factor, $\gamma_{ion}$, most studies have to take the average metallicity of the galaxy and assume some shape for its SFH.  Our conversion uses the mass, age and metallicity of each star particle in a galaxy, and therefore the true SFH of the galaxy. We calculated $\gamma_{ion}$ by modeling the SEDs of all simulated galaxies between z=4-10 with reliable star formation histories and at least 1000 stars. We modeled the SED using {\tt FSPS} \citep{fsps1, fsps2} to sum up the FUV and ionizing component from each star in a galaxy using its age, mass, and metallicity. The conversion factor $\gamma_{ion}$, shown in Figure \ref{completeness} right, is defined as

\begin{myequation}
 \int_{c/912\AA}^{c/228\AA} \ \frac{1}{h\nu} \ f_{\nu} \ d\nu
\end{myequation}

where $f_{nu}$ is the full SED of the simulated galaxy, and the integral is from 1-4 Rydbergs. From the SED, we calculated the number of ionizing photons and fit a power-law relation of ionizing emissivity as a function of $\MUV$,
\begin{equation}
    \log \gamma_{ion} = 44.55\pm(0.09) -0.473\pm(0.005) \MUV.
\end{equation}
With the updated SSP models from {\tt FSPS}, and taking the true SFH of each galaxy into account, brighter galaxies have higher ionizing emissivities compared with previous conversions which use a simple power-law SED model \citep{kuhlen12}. Our ionizing emissivity is dominated by dimmer galaxies, so this does not have a significant effect for this study, but taking the full SFH and metallicity into account is important.

\subsection{Escape Fractions}
Escape fractions of high redshift galaxies, which possibly drive reionization, are still very unknown. High redshift measurements of $\fesc$ are impossible due to the high optical depth of intergalactic neutral hydrogen between us and the sources, so any escaping ionizing radiation would be absorbed before reaching us. Measurements of $\fesc$ for lower redshift galaxies are small, ranging from 1\%-10\%, and most observations are just upperlimits. \citep{leitherer95,bland_hawthorn99,heckman01,grimes07}. Low $\fesc$  values are troubling for models trying to form a consistent view of reionization since the UV luminosity density, and therefore intrinsic ionizing emissivity, decreases at high redshifts. In order for the decreasing galaxy ionizing emissivity to have enough escaping radiation to reionize the universe, galaxies at high redshift must have had higher $\fesc$ values. So even though local, observed values for escape fractions are low and high redshift values are uncertain, most reionization models must paint high escape fractions uniformly onto all galaxies, a common value being ($\sim$ 30\%) at high redshifts \citep{finkelstein14,HM12}. Some models also allow for a variation in redshift\citep{HM12}, where higher redshift galaxies have higher $\fesc$.

Reionization models with high escape fractions for all galaxies might seem ambitious, but there are very few constraints on $\fesc$ for fainter galaxies. These galaxies do not have much star formation individually, but as a population they can contribute significantly to reionization, especially if they have higher escape fractions. One line of reasoning for them having a higher $\fesc$ than their local, more massive counterparts is that fainter, low mass halos are characterized by bursty SFHs
\citep{tolstoy09,kauffmann14,weisz14,governato15} and shallower potential wells. Star formation events, and its proceeding SN ejecta, are contained within bursts and can therefore disrupt the gas distributions in fainter galaxies more easily and create a smaller covering fraction of neutral hydrogen. So it is encouraging that faint galaxies may have a higher $\fesc$ because the faint end slope of the luminosity function is steeper in the early universe, implying a higher density of these faint, but possibly potent, galaxies at higher redshifts.

To help constrain the contribution of galaxies to reionization, we measure $\fesc$ for the resolved galaxies in our simulation. Modeling $\fesc$ in simulations is highly dependent on the distribution of gas relative to stars, as well as a the SFH. To properly model these properites we use the resolution threshold of $10^5$ dark matter and $10^5$ sph particles to determine that a galaxy has a realistic 3D distribution of gas and stars, as studied in detail in \cite{christensen10} and discussed previously in \S3.1. For our SFHs, we have a realistic star formation + feedback model \citep{fry15} where bursts are regulated by feedback and accretion rates, and create the observed fraction of stars in bursts. This results in galaxies with photometric and kinematic properties comparable with real galaxies
\citep{governato09,oh11,christensen14,kassin14,brooks11}, where \cite{governato09} and \cite{brooks11} analyzed simulations at the same resolution as the \vulcan. These physical properties all contribute to simulating realistic SFH and therefore escape fractions.

{\it To calculate the escape fraction of ionizing photons from each galaxy, we project the neutral hydrogen column density out to the virial radius and onto the 2D sky of each star particle younger than 50 Myr old in a given halo.} Previous results have used radiative transfer post processing to simulate the effect of ionizing emission on the neutral hydrogen within a galaxy. To account for local radiative transfer effects that our simulation does not include, we cut out 350 pc HII regions around each star particle that is younger than 10 Myr old. Including HII regions accounts for the radiation effects at a given time but does not self consistently evolve the ionizing radiation field. To calculate the size of an HII region around a star particle, we accumulated the sizes of individual HII regions around the O and B stars contained in a star particle, or stellar population. Each star particle can be thought of as a stellar population because we apply an IMF to each star particle to determine physical characteristics, such as SN rates or HII region size. The initial mass of a star particle is $6.46 \times 10^4$ \Msun, and assuming a Kroupa 2001 IMF, each star particle represents $\sim 40$ O stars and $\sim 150$ B stars. In comparison, a Salpeter IMF would create $\sim 25$ O stars and $\sim 100$ B stars. Assuming an O star has an HII region of $\sim$ 100 pc and B stars have HII regions of $\sim$ 10 pc \citep{stromgren39}, this averages to a star particle HII region of $\sim$ 350 pc, just at our resolution limit. Assuming a Salpeter IMF, with fewer O and B stars, would lead to smaller HII regions.

{\it Our tests show that the escape fraction is not sensitive to HII region size}. We measured $\fesc$ varying the HII region size from 0-750 pc and saw very little difference in $\fesc$ across this range. The HII regions make very little difference in $\fesc$ because HII regions are most potent in central, dense parts of the galaxy where many star particles are forming, and therefore the HII regions accumulate. In these dense regions the escape fraction is so low, due to the high column densities of neutral hydrogen, that the escape fractions of star particles change from $\sim 10^{-8}$ to $\sim 10^{-5}$, and therefore have no measurable effect of increasing the escape fraction. For young stellar populations closer to the edges of the faint galaxies, where stellar and gas densities are low, HII regions are not very effective either because there is not a clustering of stellar particles for HII regions to accumulate. For stellar populations closer to the edges of faint galaxies, after $\sim 10$ Myrs, when SNe have gone off, the stellar populations blister to the surface and dramatically increase the escape fraction, regardless of HII region size.

\begin{figure*}
 \minipage{\textwidth}
\includegraphics[trim=5mm 0mm 0mm 0mm, clip=true, width=17.5cm]{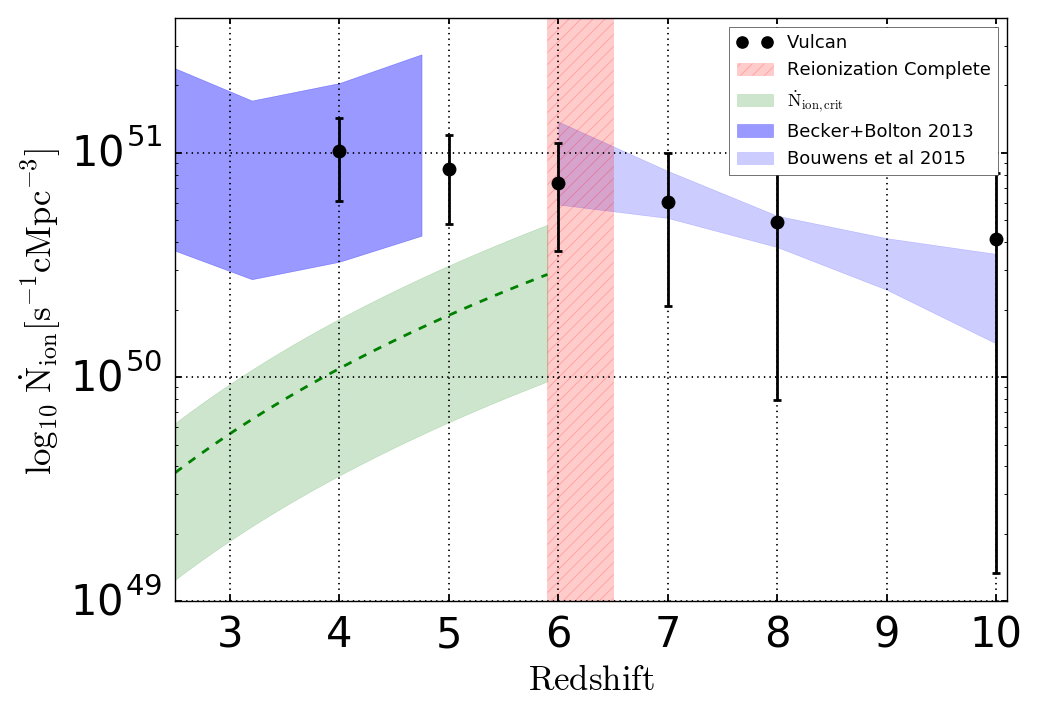}
\endminipage
\caption{~Evolution of ionizing emissivity: The black points are the ionizing emissivity calculated by convolving the simulated UVLF with the simulated relationship of $\gamma_{ion}(\MUV)$ and $\fesc(\MUV)$, and the error bars are the $1\sigma$ confidence interval from propagating the simulated distributions for the UVLF parameters and $\fesc(\MUV)$ using a Monte Carlo method. The evolution agrees well with constraints from various observations shown in blue. The red band represents when reionization is complete, and the green band represents the minimum ionizing emissivity required to keep the universe ionized once it has been reionized. The simulated ionizing emissivity agrees with observations, and is above the threshold to keep the universe ionized post reionization. The evolution is very flat because it is dominated by galaxies dimmer than $\MUV \sim -17$, whose number density is not evolving significantly, shown both in our simulations and observations. The observed ionizing emissivity can be accounted for during reionization, as well as after up to $z\sim4$, by our simulated galaxies alone, with high number densities of faint galaxies with high escape fractions.}
\label{nIon}
\end{figure*}

{\it In contrast, the escape fraction is more sensitive to the stellar age threshold for stars included in the $\fesc$ calculation}. We measured $\fesc$ varying the age threshold for stars included in the calculation from 10-50 Myrs old and saw a more noticeable difference in $\fesc$. Only including stars younger than 10 Myrs old decreases $\fesc$ by more than an order of magnitude, compared with including 25 Myr old or 50 Myr old populations, similar to results seen in \cite{ma15}. The youngest stellar populations have very high ionizing emissivities, but very low $\fesc$ because they are still embedded in their birth cloud. In comparison, a 25 Myr old stellar populations has an ionizing emissivity that is 10x lower compared with a 10 Myr old stellar population, but in our simulations, on average, the 25 Myr old stellar populations have a escape fraction that are $\sim 30$ times higher, especially at high redshifts, making them a more potent contributor to escaping ionizing radiation. Similarly, a 50 Myr old stellar population has an escape fraction comparable to the 25 Myr old stellar population, on average, but an ionizing emissivity that is 10x lower than the 25 Myr old population. So, for our calculation, we chose a threshold of 50 Myrs old.

The measured escape fraction for our simulated galaxies is the average open sky fraction of all young star particles in a halo, given by

\begin{subequations}
\begin{align}
\tau(\theta, \phi) &= \int_0^{R}n_{H}(r, \theta, \phi) \sigma_{H,912 \AA}dr \\
\fesc &=  \Bigg\langle \frac{1}{4\pi} \int e^{-\tau(\theta, \phi)} d\Omega \Bigg\rangle_{N_*}
\end{align}
\label{eqfesc}
\end{subequations}
where $\tau$ is the optical depth to ionizing radiation, $n_H(r, \theta, \phi)$ is the neutral hydrogen number density within a halo, and $\sigma_{H,912 \AA}$ is the cross section for hydrogen absorption at 912 \AA\ where absorption is strongest. Equation \ref{eqfesc}a, the integral for $\tau$, is the neutral hydrogen column density out to R, the virial radius, for a stellar particle. In equation \ref{eqfesc}b, the exp$[-\tau]$ is the probability of an ionizing photon reaching the IGM along a given line of sight. Integrating this probability over the sky, d$\Omega$, gives the average open sky fraction for an individual young star particle. We then take the average open sky fraction of all young stars in the halo, $N_*$, represented by the average brackets. This represents $\fesc$ for a halo, and our simulated measurements are shown in Figure \ref{efMuv} as a function of $\MUV$.

The escape fraction for each halo is dominated by a few star forming regions having very high escape fractions, while most stars are completely opaque to the HI distribution within a galaxy and have an escape fraction of nearly zero. The high $\fesc$ values for individual stellar populations is due to the star particle sitting on the outskirts of the HI distribution of the galaxy. These stellar populations will go through episodes of high $\fesc$ after stellar feedback (i.e. SNe) disrupts the surrounding gas distribution and the stellar population blisters to the surface. Then after 10s of millions of years the stellar population becomes dynamically enshrouded by neighboring gas, and $\fesc$ for that star particle drops to a low value again. The blistering of stellar populations is more likely in bursts of star formation, so less bursty galaxies, as well as less bursty star formation prescriptions, will have a lower $\fesc$. Our star formation prescription has been shown to be bursty, especially in dwarf galaxies \citep{fry15}, consistent with real galaxies \citep{kauffmann14}, and consistent with our fainter galaxies have a higher $\fesc$. The central, dense regions of galaxies containing cold, molecular gas is currently difficult to realistically simulate at all resolutions, but we have shown that our star formation prescription creates realistic lower density, warm and hot ISM \citep{governato07,ohsim11}. So we simulate a realistic ISM where the escape fractions are measurable and significant.

To encapsulate the escape fractions in the total ionizing budget from galaxies, we fit the trend of our simulated $\fesc$ as a function of $\MUV$ for each halo. In this form we can convolve it with the UVLF to calculate the ionizing photon budget from galaxies. Figure \ref{efMuv} shows our measured values of both $\fesc$ and $\MUV$ for the resolved halos in our simulation, spanning $z\sim4-8$. At $z\sim10$, none of the halos have passed our resolution threshold. Faint galaxies tend to have higher escape fractions, $\fesc>1\%$, than brighter galaxies, though the brighter galaxies have more scatter. The smaller scatter for faint galaxies relative to the brighter galaxies may be due to fewer faint galaxies passing our resolution threshold, but not many bright galaxies have $\fesc>0.1\%$. The blue points are observed $\fesc$ values from local galaxies, and the red points are observed upper limits from local galaxies \citep[and citations therein,Choi+16]{leitet13}. The observed galaxies also show a large scatter in $\fesc$. These galaxies with measurable escape fractions were targeted for their high specific SFR and therefore likely to have a measurable, and high, escape fraction. Our simulated, brighter halos scatter up to $\sim 1\%$, but most lie below the observed, upper limit constraints. Here we have fit a power law to the simulated measurements using the maximum likelihood technique, with the equation
\begin{equation}
\mathrm{log \ \fesc = (0.51 \pm 0.04) \MUV + 7.3 \pm 0.8}
\end{equation}
up to a maximum escape fraction of $\sim 35\%$, the maximum $\fesc$ that the linear fit is sampled by a measured halo at $\MUV \sim -15$.
It is difficult to quantify uncertainties in simulations, so here we compute the variance in the relationship via the posterior distribution of the power law fit to the relationship, assuming each individual $\fesc$ and $\MUV$ value has no error.
Due to the resolution of the simulation, the bright portion of the plot is well sampled by galaxies, but as you move to fainter magnitudes there are fewer galaxies with resolved morphologies.
This leads to the faint portion of the relationship, and perhaps the more interesting portion, being constrained by $< 10$ simulated galaxies.
This leads to the large uncertainties for the escape fractions of faint halos.
In future work, we plan to increase the sample size by running a similar volume size but at 10x the particle resolution, so we will increase our sample size at these faint magnitudes by about a factor of 10 as well.
These findings are rather high relative to previous results of simulated escape fractions \protect\citep{ma15}, and this might be a resolution effect.
We ran a convergence test, see Appendix, but at this resolution we still aren't resolving the scales of giant molecular clouds and other interesting scales within the ISM that are important.
By not resolving the more complicated structures, the chimneys that our stellar feedback blows into our gas distribution might be unphysically large.
The results from the future, higher resolution simulation will also help alleviate this issue.

We followed the same procedure but fitting $\fesc$ as function of SFR instead of $\MUV$. The SFR and $\MUV$ are highly correlated so the relationship looks very similar, with a sign flip and slightly steeper slope

\begin{equation}
\mathrm{log \ \fesc = (0.70 \pm 0.09) log \ SFR - 2.5 \pm 0.1}
\end{equation}

Here, galaxies with SFR $\sim 0.01 M_{\odot}$/yr have $\fesc \sim 10s\%$. This relationship represents {\it brighter galaxies having little to no escaping ionizing radiation}, and supports the model where {\it faint, low mass galaxies contribute a significant fraction of their intrinsic ionizing radiation to reionization}. To calculate the total number of ionizing photons in our volume, we integrate down to $\MUV \sim -14$ to $-15$, depending on redshift and completeness level, but not more than a magnitude past the constrained portion of our fit.

\section{Results: Consistent View of Reionization}

\begin{figure*}
\centering
 \minipage{0.5\textwidth}
  \includegraphics[trim=0mm 0mm 0mm 0mm, clip=true, width=8.75cm]{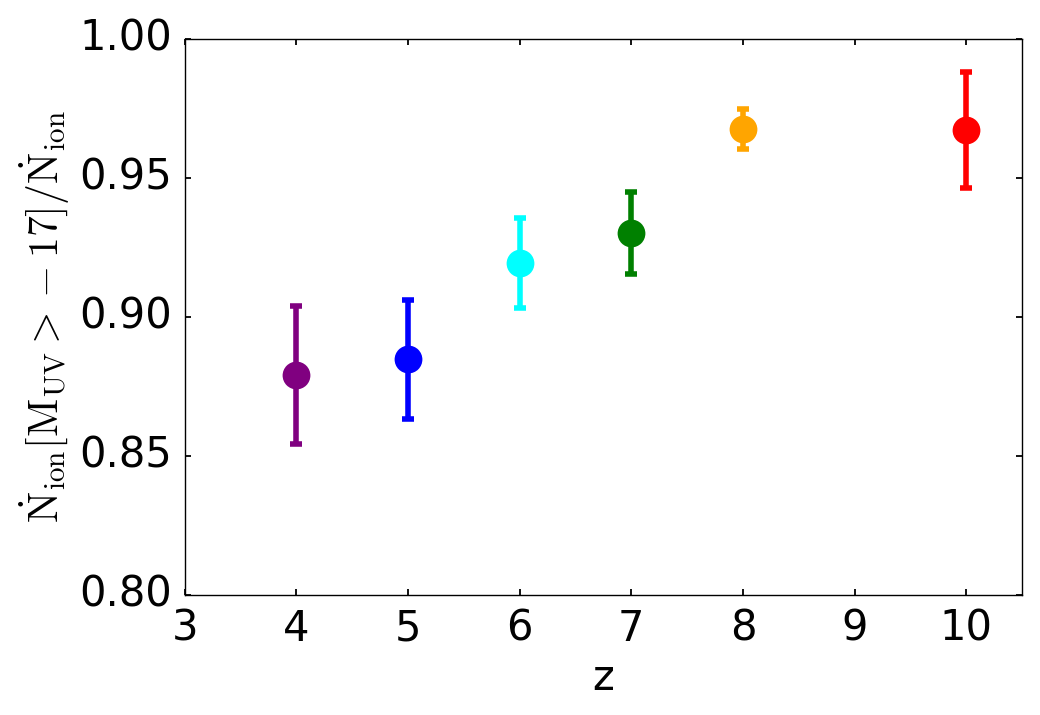}
  \endminipage
  \minipage{0.5\textwidth}
  \includegraphics[trim=0mm 0mm 0mm 0mm, clip=true, width=8.75cm]{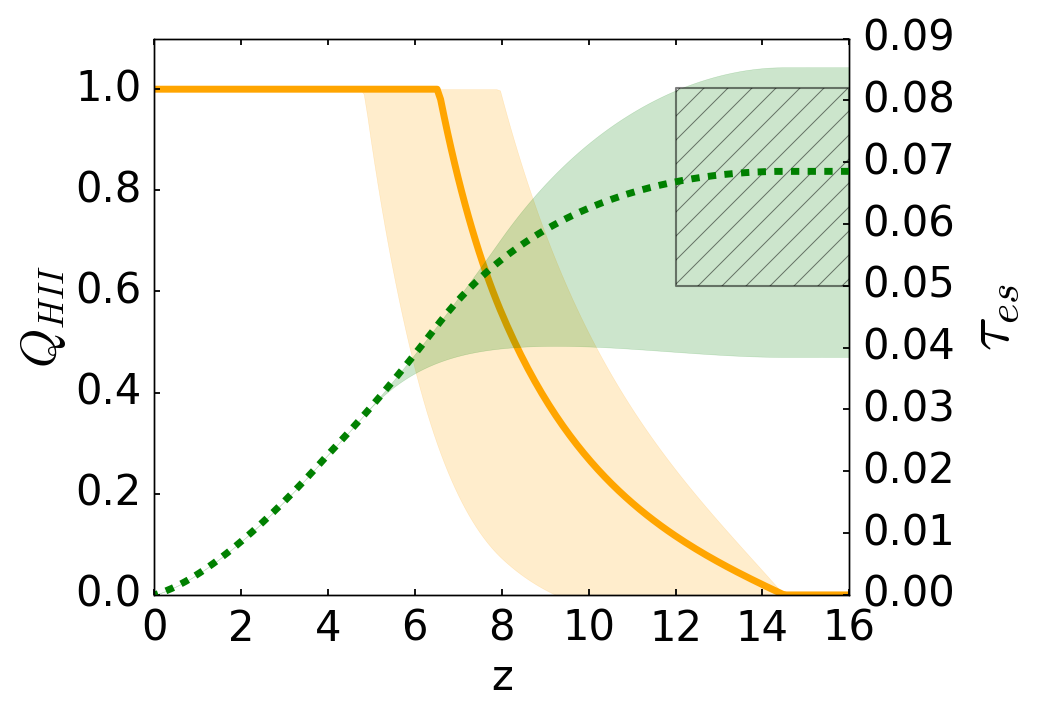}
  \endminipage
\caption{{\sc Left:} ~Evolution of contribution of faint galaxies to the ionizing emissivity: Faint galaxies, defined as being fainter than $\MUV = -17$, the observational limit at high redshifts, dominate the contribution of escaping ionizing photons to the ionizing emissivity. At high redshifts their contribution is 98\% and by $z\sim4$ is still at 88\%. {\sc Right:} ~Evolution of the fraction of ionized hydrogen from our simulated ionizing emissivity in orange (left axis), and the integrated optical depth to the CMB from this evolution in green (right axis). The error bars for each are the $1\sigma$ confidence interval from propagating the ionizing emissivity distribution using a Monte Carlo method. The hatched region represents constraints on the optical depth from \protect\cite{planck15}. Using the ionizing emissivity shown in Figure \ref{nIon}, which is dominated by the faint galaxies, we reach 50\% ionization at $z \sim 8$, and complete reionization by $z \sim 6.5$. We have a more extended reionization history than that inferred from recent measurements of the UVLF, which increases our optical depth measurement to be in agreement with the most recent CMB measurements.}
\label{tau}
\end{figure*}

The primary uncertainties in the contribution of faint galaxies to reionization are their number densities, quantified by the faint end slope of the luminosity function, and their escape fractions. The number densities of high redshift, low luminosity galaxies are becoming more constrained with deeper surveys that can probe to fainter magnitudes at high redshift. Even though surveys are observing to fainter limits, reionization models must still project the observed luminosity functions to fainter magnitudes than they observe to accumulate enough photons for reionization. So faint galaxies, which may be numerous at high redshift, are difficult to observe, but escape fractions of high redshift galaxies are impossible to observe due to absorption from intergalactic neutral hydrogen between us and the source galaxy. To approach this problem from a theoretical perspective, we use a high resolution, uniform volume simulation to constrain the UVLF to $\sim$ two magnitudes fainter than observations, and use the detailed 3D distribution of gas and stars in resolved high redshift galaxies to calculate $\fesc$ as a function of $\MUV$. With these findings we now constrain the ionizing emissivity from galaxies during reionization in our simulation, and compare with observed constraints from the lyman alpha forest and cosmic microwave background.

The comoving ionizing emissivity of galaxies in the \vulcan is calculated by integrating the galaxy UVLF, convolved with the conversion from $\MUV$ to ionizing emissivity $\gamma_{ion}$, and the escape fraction $\fesc$ as a function of $\MUV$.
\begin{equation}
    \dot{N} = \mathlarger{\int_{-\infty}^{M_{UV,lim}}d\MUV \phi(\MUV)\gamma_{ion}(\MUV)\fesc(\MUV)}
\end{equation}
where $M_{UV,lim}$ is our 50\% completeness limit at a given redshift, varying between $\MUV\sim-14$ to $-15$ depending on redshift. Specifically the faint limit is [-13.9, -14.3, -14.9, -15.0,  -15.6, -15.3] for z = 4, 5, 6, 7, 8, 10 respectively. We chose the 50\% completeness limit, rather than integrating all the way down our completeness correction to avoid becoming completeness dominated. With a rising mass function towards lower mass halos the completeness correction becomes highly uncertain. This integral is the ionizing emissivity that escapes galaxy halos and contributes to reionization. The steep slope of the UVLF at the faint end convolved with faint galaxies having high escape fractions, $\fesc\sim35\%$, creates a high ionizing emissivity, as shown in Figure \ref{nIon}. The black points are from the \vulcan, with the error bars representing the $1\sigma$ uncertainties from propagating the uncertainties of the simulated UVLF parameters and $\fesc(\MUV)$ via Monte Carlo sampling of the posteriors.

The comoving ionizing emissivity in the real universe cannot be directly measured, but it can be inferred from various observational measurements of the high redshift universe, shown as the blue regions in Figure \ref{nIon}. \cite{becker13} constrained the ionizing emissivity at $z\sim2-5$ using measurements of the lyman alpha forest; they combined measurements of the photoionization rate and the mean free path of ionizing photons, shown as the darker blue region. The lighter blue region, from \cite{bouwens15}, sampled models of the ionizing emissivity and its evolution to determine which models were compatible with various observations: the optical depth to the surface of last scattering from Planck \citep{planck}, the ionized hydrogen fraction at various redshifts \citep{schenker14}, the completion of reionization by z $\sim$ 6 \citep{fan06}, and the inferred ionizing emissivity from \cite{becker13}. The comoving ionizing emissivity of galaxies in the \vulcan, calculated by integrating the galaxy UVLF using equation 5, has large uncertainties but shows excellent agreement with the observationally constrained ionizing emissivity. Our simulated galaxies alone, with a high number density of faint galaxies, and those faint galaxies having high escape fractions, can account for the ionizing emissivity responsible for reionizing the IGM.

Another constraint is that the ionizing emissivity must also keep the universe ionized after reionization. As galaxies spew ionizing radiation into the IGM, reionization continues until $z \sim 5.9 - 6.5$, when measurements of the Gunn-Peterson optical depth imply that reionization is complete \citep{fan06}, the red hashed region in Figure \ref{nIon}. Past this point, once reionization is finished, there is a critical minimum number of ionizing photons required to keep the universe ionized. That is, once the universe is in ionization equilibrium, the ionizing emissivity must balance the recombination rate of the fully ionized IGM given by
\begin{equation}
\dot{N}_{ion}^{crit} = C_{HI}\alpha_A(T_0)n_H(1+Y/4X)(1+z)^3
\end{equation}
where $C_{HI}$ is the neutral hydrogen clumping factor, $\alpha_A$ is the case A recombination coefficient for hydrogen, and $T_0$ is the IGM temperature. Given fiducial values, taken from \cite{kuhlen12}, the critical ionizing emissivity is
\begin{equation}
\dot{N}_{ion}^{crit} = 3 \times 10^{50} s^{-1} cMpc^{-3} \left(\frac{C_{HI}}{3}\right) \left( \frac{T_0}{2 \times 10^4 K}\right)^{-0.7} \left( \frac{1+z}{7} \right) ^3
\end{equation}
This is shown as the green shaded region in Figure \ref{nIon}, assuming an IGM temperature of $2\times10^4$ K, and the width is varying the clumping factor within the uncertainties of 2-5. Early simulations concluded that the clumping factor was high \citep{gnedin97}, but these early simulations included the clumpiness of neutral hydrogen within halos. The clumpiness within halos is now taken into account using the separate quantity $\fesc$, so more recent determinations of the clumping factor in the IGM are lower. More recent simulations have constrained the clumping factor to lie between 1-5 \citep{pawlik09,shull12,finlator12}, and most groups assume a value of 3. So post reionzation, at $z < 6$, a realistic galaxy ionizing emissivity should be above $\dot{N}_{ion}^{crit}$. Figure \ref{nIon} shows that {\it our simulated measurements follow the expected evolution of the ionizing emissivity} from \cite{bouwens15} and \cite{becker13}, and is above $\dot{N}_{ion,crit}$ at $z>6.5$, suggesting that {\it galaxies are the sources that drive the reionization of the universe, with a large number density of faint galaxies that have high $\fesc$.}

The evolution of our ionizing emissivity is very flat because it is dominated by faint galaxies at all redshifts, whose number densities are not evolving very much. The faint end slope is becoming shallower with time, but the over all normalization is rising with time, creating a consistent number density of faint galaxies from $z\sim4-10$. This trend is seen in both observations \citep{bouwens15b} and our simulations. This consistent, high number density of faint galaxies, convolved with their high escape fractions, makes galaxies with $M_{UV} > -17$ dominate the ionizing emissivity, as shown in Figure \ref{tau}, left panel. The left panel of Figure \ref{tau} shows the fraction of the simulated ionizing emissivity coming from galaxies fainter than $M_{UV} = -17$ as a function of redshift. At $z \sim 10$ they contribute $\sim$ 99\%, and at $z \sim 4$ they still contribute $\sim$ 88\%. So throughout reionization they are the dominant contributor, but even post reionization they still have a significant contribution to the background ionizing emissivity keeping the IGM ionized. Each faint galaxy individually doesn't have much star formation, but as a population they have a significant budget of escaping ionizing photons.

The evolution of the ionizing emissivity, as well as the expansion of the universe, dictate the evolution of ionized hydrogen during reionization. This is quantified as the volume filling fraction of ionized hydrogen $Q_{HII}$ and its evolution is constrained by the differential equation
\begin{equation}
    \frac{dQ_{HII}}{dt} = \frac{\dot{N}_{ion}}{\bar{n}_H} - \frac{Q_{HII}}{\bar{t}_{rec}},
\end{equation}
which is the balance of ionizing events and recombination events. Here $\dot{N}_{ion}$ is our measured comoving emissivity from Figure \ref{nIon}, $\bar{n}_H$ is the mean comoving hydrogen number density, and $\bar{t}_{rec}$ is the mean recombination rate in the IGM. We linearly interpolate our simulated measurements of the emissivity, and extrapolate to $z\sim14.5$. Solving this ordinary differential equation, given our evolving emissivity, we get the evolution of $Q_{HII}$ shown in Figure \ref{tau}, right panel. The orange line represents the average history of ionized hydrogen given by the average emissivity measurements, and the shaded regions encapsulate the $1\sigma$ uncertainties in the emissivity evolution shown by the error bars in Figure \ref{nIon}. The \vulcan is 50\% reionized by $z \sim 8$ and completely reionized by $z \sim 6.5$, in excellent agreement with observations of the Lyman alpha forest \citep{fan06} and the evolving visibility of Lyman alpha emission from galaxies \citep{schenker14}. Our reionization is more extended than recent measurements from the UVLF, starting at a redshift of 14.5. We did not measure the UVLF beyond $z \sim 10$, but allowed our emissivity to be extrapolated to $z \sim 14.5$ because our CSFR is flat, as seen in Figure \ref{paramsearch}, right. A flat CSFR implies an equally significant amount of SF is going on from $z\sim10-14.5$, so variations in the ionizing emissivity coming from young stars are minimal.

The other strong observational constraint on reionization is the optical depth to the surface of last scattering. Given $Q_{HII}(z)$, the fraction of ionized hydrogen as a function of redshift, we can calculate the expected optical depth of CMB photons from electron scattering given by the integral
\begin{equation}
    \tau_e = \int_{0}^{\infty} dz \frac{c(1+z)^2}{H(z)} Q_{HII}(z)\sigma_T\bar{n}_{H}(1 + \eta(z) Y/4X)
\label{eqtau}
\end{equation}
where $\sigma_T$ is the Thomson scattering cross section, and $\eta$ is the ionization state of He, given by 1 for singly ionized and 2 for doubly ionized. We assume that He is singly ionized for $z > 4$ and doubly ionized there after. By integrating Equation \ref{eqtau} we get $\tau_{es}  = 0.07_{-0.03}^{+0.02}$, shown as the green region in Figure \ref{tau}, right panel, and in excellent agreement with \cite{planck15}, shown as the hatched region representing $\tau_{es}  = 0.066 \pm 0.016$. The CMB measurement is a strong constraint, but an integrated quantity, so the agreement validates our integrated $Q_{HII}$ but not our history of reionization. The optical depth is very sensitive to $Q_{HII}$ at high redshifts, when the physical number densities were much higher. A small number of free electrons at high redshift increases $\tau_{es}$ more than an equal number at lower redshifts. So taking SF in the early simulation into account and extrapolating the emissivity out to $z\sim14.5$, instead of to $z\sim10$, a difference of $\sim 200$ Myr, increases $\tau_{es}$ by $\sim20\%$.

\section{Conclusions}
The sources that reionized the universe are still unknown, but the ionizing photon budget from faint galaxies, below the detection limit of current surveys, may be significant. Observations point to a high number density of faint galaxies at high redshift, and that these faint galaxies, with bursy SFH, may have high escape fractions for ionizing radiation. Here we approached the problem from the theoretical perspective and used a large, uniform volume cosmological simulation, the \vulcan, to constrain the shape of the UVLF and measure $\fesc$ for a statistical sample of resolved systems from $z \sim 4-10$. We optimized the subgrid SF parameters to create realistic galaxies at $z=0$, and are therefore predictive at $z \sim 4-10$, during reionization. The simulation includes a large population of galaxies with realistic SFHs and resolved morphologies, giving us a statistical sample of galaxies to constrain the shape of the UVLF and measure escape fractions. In our simulations, we constrain the shape of the UVLF down to $M_{UV} \sim -15$, two magnitudes fainter than observations, and measure a steeper faint end slope, and therefore a relatively higher number density of faint galaxies, during reionization. We test the convergence of our $\fesc$ measurements at the resolution of the \vulcan, and measure $\fesc$ for the systems passing our strict resolution requirements, down to $\MUV \sim -15$. We find a strong correlation of $\fesc$ with $\MUV$, where bright systems have $\fesc < 0.1\%$ and faint systems have $\fesc \sim 35\%$. Convolving these findings, we calculate an ionizing emissivity from galaxies that is consistent with observations from $z \sim 4-10$. This simulated emissivity reionizes the universe at the proper rate and completes reionization at $z\sim6.5$. With this reionization history, the optical depth to the surface of last scattering is $\tau_{es}  = 0.07_{-0.03}^{+0.02}$, in excellent agreement with the most recent Plank results. Therefore, we conclude that faint galaxies as a population, with high number densities and high escape fractions, have the proper ionizing photon budget to reionize the universe in agreement with constraints from observations.

\begin{figure*}
 \minipage{0.5\textwidth}
  \includegraphics[trim=0mm 0mm 0mm 0mm, clip=true, width=9.0cm]{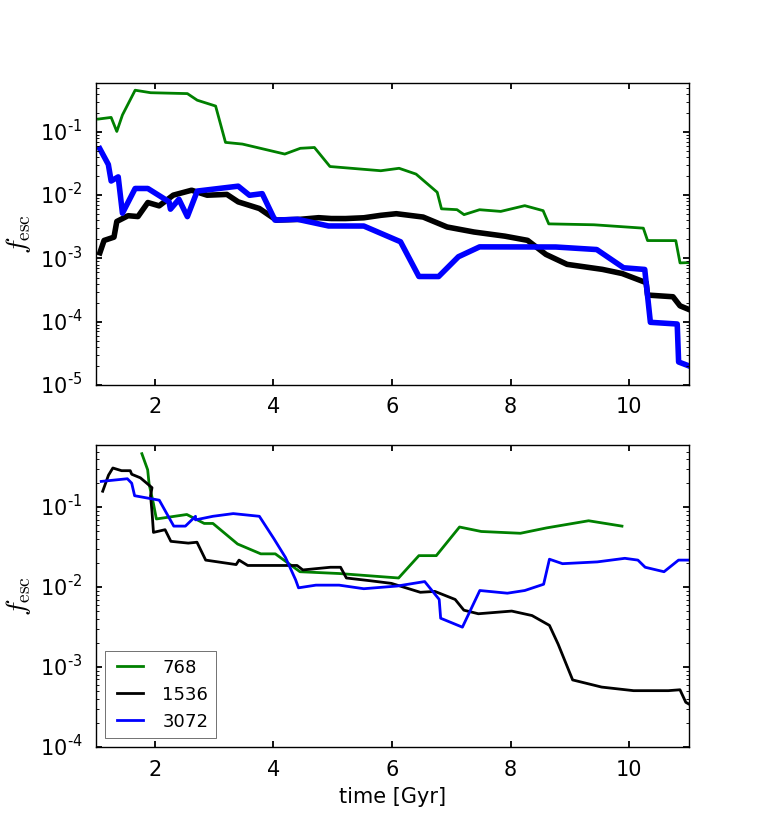}
  \endminipage
  \minipage{0.5\textwidth}
  \includegraphics[trim=0mm 0mm 0mm 0mm, clip=true, width=9.0cm]{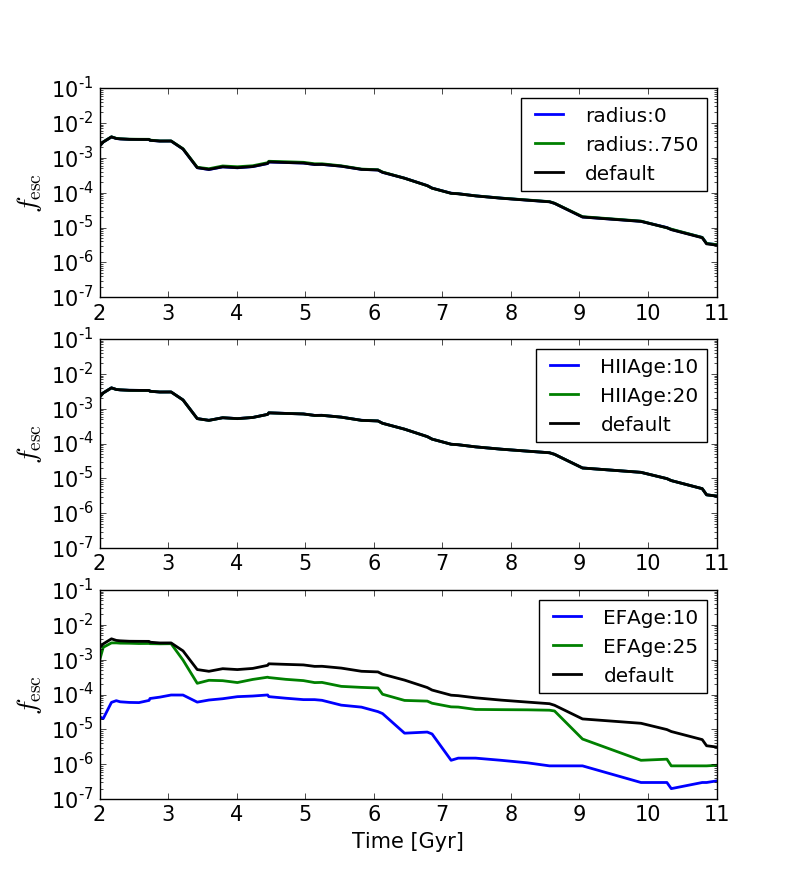}
  \endminipage
\caption{{\sc Left: Convergence of $\fesc$} ~The evolving escape fraction, averaged over 10 steps, for a $\sim 10^{11}$ \Msun\ halo (top), and  its most massive subhalo (bottom), each ran at 3 resolutions: super-\vulcan (blue), \vulcan (black), and sub-\vulcan (green). The super-\vulcan and \vulcan escape fractions, shown as thicker lines, converge to a similar measurement once both have crossed our resolution threshold at t = 2 Gyr. Here they both have accreted at least $10^5$ dark matter particles and $10^5$ sph particles. For the subhalo (bottom), the \vulcan and sub-\vulcan simulations never cross our resolution threshold and therefore the measured $\fesc$ does not converge. In conclusion, having a strict resolution threshold is important for claiming converged $\fesc$ values. {\sc Right: $\fesc$ Parameter Test:} ~The evolving escape fraction, averaged over 10 steps for a $\sim 10^{11}$ \Msun\ halo ran at the \vulcan resolution, varying our parameter choices associated with $\fesc$ in each panel. The default parameters are 350 pc HII regions around stellar populations younger than 10 Myr old, and a threshold stellar age of 50 Myrs. Varying parameters associated with the HII regions, their size (top) and age (middle) does not affect $\fesc$ significantly. However, varying the stellar age threshold for star particles included in the calculation (bottom) can change $\fesc$ by more than an order of magnitude. 10 Myr old stellar populations contribute the most ionizing photons but have very low escape fractions. 25-50 Myr old stellar populations contribute $\sim10-100\times$ fewer ionizing photons, but have $\sim50\times$ higher $\fesc$ making them more potent contributors.}
\label{resTest}
\end{figure*}

\section{Appendix}

{\bf Convergence Test:} We used the criteria from \cite{christensen10} to define when a galaxies morphology is resolved and therefore the escape fraction is believable. \cite{christensen10} investigated the resolution sufficient to establish galactic SFRs, SFHs, stellar feedback, and the distribution of gas and stars. They show that there is a minimum number of particles required to create realistic star formation histories (SFH), and a higher threshold for realistic morphologies. Insufficient resolution contributes to unphysically altering positions of the particles and the SFR of the galaxy. Correctly determining the amount and location of SF is vital in simulated galaxies because SF produces and distributes metals throughout the galaxy, affects the distribution of matter in the galaxy through feedback, and enables us to relate simulations to observations. In particular, modeling $\fesc$ in simulations is sensitive to the SFH and the 3D distribution of gas and stars, or the number of young stars and their positions with respect to neutral hydrogen. \cite{christensen10} found that to properly model the SFH and morphology of a galaxy, a halo must have $10^5$ dark matter and $10^5$ sph particles.  to determine that a galaxy has a realistic 3D distribution of gas and stars, as studied in detail in \cite{christensen10} and discussed previously in \S3.1. Below this limit, the 3D distribution of gas and stars is not accurate and will give erronous values for $\fesc$. This limit is beyond that needed for a realistic SFH, $10^4$ dark matter and $10^4$ sph particles, so by passing the morphology requirement a galaxy also has a convered SFH.

To show that a resolved morphology leads to a converged $\fesc$ measurement, we ran a zoomed in simulation, which creates a sub Milky Way mass galaxy at $z = 0$, at three resolutions: The same as the \vulcan, $8\times$ the resolution of the \vulcan (super-\vulcan), and 1/8 times the resolution of the \vulcan (sub-\vulcan). The super-\vulcan system crosses the resolution threshold at $\sim$ 2 billion years, the \vulcan system crosses at $\sim$ 4 billion years, and the sub-\vulcan system never accreted enough particles to have a resolved morphology. We measure the escape fractions following the same methodology in Section 3.3, including cutting out HII regions around young star particles. The escape fractions are constantly fluctuating, following the stocasticity of the star formation rate, and its subsequent SNe blowing holes in the HI distribution; so for a better comparison of the escape fraction over time, we averaged the escape fraction of the halo over the neighboring 10 steps. For the main halo progenitor, Figure \ref{resTest}, left top, after $\sim$ 2-4 billion years, the super-\vulcan and \vulcan escape fractions, shown as the thicker lines, converge on a measurement of the escape fraction, with differences mostly being due to variations in the star formation histories due to stochasticity. The most massive subhalo in the zoomed in simulation also shows that having a strict resolution threshold is important for a converged $\fesc$ measurement. For the most massive subhalo Figure \ref{resTest}, left bottom, only the super \vulcan simulation reaches our resolution threshold in the age of the universe, after $\sim$ 4 billion years, and the escape fractions do not converge. Also, for the super \vulcan simulation, $\fesc$ is higher at higher redshifts, and higher for lower mass systems, in agreement with theory, including our conclusions.

Having a strict resolution threshold of $10^5$ dark matter particles and $10^5$ sph particles is important for claiming converged results for $\fesc$. Our tests show that above this threshold $\fesc$ values converge, but below this threshold they do not.

{\bf $\fesc$ parameter test:} When we calculate the covering fraction of neutral hydrogen for a halo we chose certain parameters, so here we test if the escape fraction calculation is sensitive to these choices of parameters using the same \vulcan resolution, zoomed in run for the convergence test. We varied the sizes of the HII regions we cut out around young stars from 0 to 750 pc, the maximum age of an HII region from 10-20 Myrs, and the maximum stellar age that we used in our covering fraction measurement from 10-50 Myrs.

{\it HII Regions}: Before we measure the escape fraction we first take the local radiation field of the stars into account and cut out HII regions around stars younger than 10 Myrs old that are 350 pc in size. Increasing the size of the HII region, as well as how long it lives, decreases the amount of neutral hydrogen around young stars, especially in high density regions where many young stars cluster. In high density regions, overlapping HII regions can significantly increase the mean free path of an ionizing photon. However, {\it our tests show that varying the parameters from the HII regions has very little effect on the escape fraction}. Star particles with low escape fractions live in the very central, dense regions of a galaxy where there is a lot of star formation. Here star particles are fairly clustered and have overlapping HII regions, but they are still deeply embedded in neutral hydrogen. So depleting some neutral hydrogen with the clustered HII regions increases the mean free path of an ionizing photon, but not significantly, and not to the edge of the halo. Here the HII regions are extracting the most neutral gas, but still not bring the escape fraction up significantly for those star particles. For stars that have high escape fractions, they live in lower density regions and tend to be older, greater than 10 Myrs, so they have already dispersed the gas around them to scales larger than the HII regions through SN. In these lower density regions no significant amount of gas is extracted with an HII region, so with or without the HII regions the escape fraction is the same.

{\it Stellar Ages}: Our escape fraction measurement is the average sky covering fraction of neutral hydrogen around young stars up to a threshold age. For the paper, we used a threshold age of 50 Myrs. Decreasing the maximum age of the stellar population that is included in the escape fraction calculation decreases the number of stars that are sampled to only the youngest stars. These stars tend to live in denser environments before stellar feedback can disrupt the surrounding gas of the stellar birth cloud. As you increase the threshold age of stars, you increase the sample size but older stars are not contributing as many ionizing photons as their younger counterparts since some of the more massive stars have already turned off the main sequence. Here we tested the change in $\fesc$ by varying the threshold age from 10 - 50 Myrs. Using \cite{fsps1} and assuming a Kroupa IMF and a metallicy of $0.01 Z_{\odot}$, a 10 Myr stellar population has an ionzing emissivity 10x greater than a 25 Myr stellar population, and 100x greater than a 50 Myr stellar population.

In our test simulation, decreasing the maximum age of a stellar particle included in the escape fraction calculation changes the measured escape fraction by an order of magnitude. Setting the threshold age to 50 Myr or 25 Myr gives similar results at high redshift, but having a strict threshold of 10 Myr drops the measured escape fraction by an order of magnitude. With a threshold of 10 Myrs the calculation is only sampling the stars that are still deeply enveloped by their birth cloud. So they individually have large numbers of ionizing photons but low escape fractions which average out to the halo having a low escape fraction. For this simulation, 25 Myr stellar populations on average have escape fractions that are 25x higher than 10 Myr stellar populations. So although 25 Myr old stellar populations are creating fewer ionizing photons compared with 10 Myr old stellar populations, they have significantly higher escape fractions, and are therefore more potent contributors to the ionzing flux entering the IGM. Their average escape fractions are higher because after 10 Myr SNe feedback is disrupting the birth cloud allowing portions of the sky to have a very small optical depth to ionizing radiation.

Adding HII regions is physically motivated but unnecessary at this resolution. The addition of HII regions, up to 750 pc in size, has a negligable effect on $\fesc$. {\it Dropping the age threshold to 25 Myrs will have a negligible affect on the average measured escape fraction for a halo, but decreasing to 10 Myrs will drop it significantly, and this would exclude some of the more potent contributors to reionization}. So our parameter choices of HII regions that are 350 pc in size for 10 Myrs, and a stellar age threshold of 50 Myrs are well motivated.

\section{Acknowledgements}
Fabio Governato, Tom Quinn, and Lauren Anderson were partially supported by NSF award AST-1311956 and HST award AR-13264. This research is part of the Blue Waters sustained-petascale computing project, which is supported by the National Science Foundation (awards OCI-0725070 and ACI-1238993) and the state of Illinois. Blue Waters is a joint effort of the University of Illinois at Urbana-Champaign and its National Center for Supercomputing Applications. This work is also part of a PRAC allocation support by the National Science Foundation (award number OCI-1144357). Lauren Anderson would like to thank Will Shown and Matt Wilde for helpful discussions about Photoshop, and Yusra AlSayyad and Yumi Choi for helpful discussions on reionizaiton and, in general, their moral support to finish this paper.
\clearpage
\bibliographystyle{mnras}
\bibliography{lmanders}
\bsp	
\end{document}